\documentclass{aa}

\usepackage{graphicx,float}
\usepackage{amsmath}

\usepackage{txfonts}
\usepackage{graphicx}	
\usepackage{amsmath}	

\usepackage{multirow}
\usepackage{subfigure}
\usepackage{placeins}
\usepackage{adjustbox}
\usepackage{xcolor}
\usepackage{float}
\usepackage{amsmath}	
\usepackage{lscape} 
\usepackage{caption}
\usepackage[switch]{lineno}
\usepackage{natbib}
\usepackage[colorlinks=true,linkcolor=blue,allcolors=blue]{hyperref}
\usepackage{gensymb}

\begin{document} 
\newcommand{\Pinuave}{\langle \Pi_\nu \rangle}
\newcommand{\Pinuaveopt}{\langle \Pi_\nu^{\rm opt}\rangle}
\newcommand{\nupSave}{\langle \nu_{p}^{S} \rangle}
\newcommand{\Pinuord}{\Pi_\nu^{\rm ord}}
\newcommand{\Pinuaveord}{\langle\Pi_\nu^{\rm ord}\rangle}
\newcommand{\fgammacut}{f_{\gamma_{\rm cut}}}
\newcommand{\gcutminratio}{\gamma_{\rm cut}^{\rm min. ratio}}
\newcommand{\jetset}{\texttt{JetSeT}}
\newcommand{\DIFLATEXDIFFIGNORE}[1]{#1}
   \title{Modeling and statistical characterization of synchrotron multi-zone polarization in blazars}

   \subtitle{}

   \author{A. Tramacere
          \inst{1}
          }

   \institute{Department of Astronomy, University of Geneva, Ch. d’Ecogia 16, 1290 Versoix, Switzerland\\
              \email{andrea.tramacere@unige.ch}
             }

   \date{Received August 9, 2025; accepted October 5, 2025}

  \abstract
   {Multiwavelength polarimetric observations of blazars reveal complex energy-dependent polarization behavior, with a decrease in the polarization fraction from X-ray to millimeter bands and significant variability in the electric vector position angle (EVPA). These trends challenge simple single-zone synchrotron models and suggest a more intricate turbulent jet structure with multiple emission zones.}
   {This work aims to develop a statistical framework to model the energy-dependent polarization patterns observed in blazars, particularly focusing on the behavior captured by IXPE in the X-ray and RoboPol in the optical. The goal is to determine the statistical characterization of multi-zone models, in terms of the cell size distribution, and of the distribution of the physical parameters of the electron energy distribution (EED).}
   {A Monte Carlo simulation approach was employed to generate synthetic multi-zone synchrotron emission, using the \jetset~ code, from a spherical region populated by turbulent cells with randomly distributed physical properties. Simulations were run across various scenarios: from identical cells to power-law-distributed cell sizes and EEDs with different cutoff and low-energy slope distributions. The simulation results were compared with the observed IXPE and RoboPol polarization trends.}
   {Our analysis demonstrates that a purely turbulent, multi-zone model can explain the observed energy-dependent polarization patterns without requiring a correlation between the cell size and the EED parameters. The key determinant of polarization is the effective number (flux-weighted) of emitting cells, which is significantly modulated by the dispersion in cell properties, especially the EED cutoff energy, at higher frequencies, and the dispersion in the EED low-energy spectral index, at lower frequencies.}
   {Using a fractional dispersion on the EED cutoff on the order of 90\%, and a dispersion of the EED low-energy spectral index between $\approx 0.5$ and $\approx 1.5$, our model reproduces both the chromaticity of the millimiter-to-X-ray polarization trends observed in IXPE multiwavelength campaigns for high synchrotron-peaked blazars, and the optical polarization limiting envelope, observed in the RoboPol dataset.}
   \keywords{Radiation mechanisms: non-thermal -- Polarization -- galaxies: jets –- galaxies: active}

   \maketitle

\section{Introduction}
\label{sec:intro}
Blazars are a subclass of active galactic nuclei (AGNs) in which a relativistic jet, launched by the central engine, is oriented close to the observer's line of sight \citep{Blandford1978}, resulting in highly beamed, rapidly variable, and strongly polarized emission across the entire electromagnetic spectrum \citep{Urry1995}.
Their spectral energy distributions (SEDs) exhibit two principal components: a low-energy component, with power peaking from the infrared (IR) to the X-ray band, and a high-energy component, which peaks in the $\gamma$-rays. 
In the most widely accepted scenario, the low-energy component is interpreted as synchrotron (S) radiation emitted by ultrarelativistic electrons, while the high-energy bump is attributed either to inverse Compton (IC) emission in purely leptonic models \citep{Blandford1979} or to high-energy emission from ultrarelativistic protons in hadronic models \citep{Boettcher2013}.
These sources are traditionally classified based on their synchrotron peak frequency ($\nu_p^S$), ranging from low synchrotron-peaked (LSP, $\nu_p^S<10^{14} \mathrm{Hz}$) through intermediate synchrotron-peaked (ISP,  $10^{14} \mathrm{Hz}<\nu_p^S<10^{15}  \mathrm{Hz}$) to high synchrotron-peaked (HSP, $\nu_p^S>10^{15} \mathrm{Hz}$) blazars \citep{Abdo2010}.

The radiation in the S bump is linearly polarized, and the level of the fractional polarization is very low (close to zero) in the low-energy branch, at millimiter frequencies, and reaches a level of tens of percent above the peak of the S bump \citep{marscher2022}. Moreover, \cite{Angelakis2016}, using data from the RoboPol high-cadence polarization-monitoring program \citep{King2014, Pavlidou2014}, found that for a large sample of $\gamma$-ray-loud blazars, the fractional optical polarization depends on $\nu_p^S$ showing a limiting envelope, with both the polarization and its dispersion higher in LSP sources than in HSP ones.
They also demonstrated that the randomness of the optical electric vector position angle (EVPA) is higher in LSP sources compared to HSP ones.

The synchrotron emission from these sources, extending from radio to X-ray frequencies, provides a unique window into the physical conditions within relativistic jets, including magnetic field structure, particle acceleration mechanisms, and jet dynamics \citep{Marscher2014}. Recent observational advances have dramatically enhanced our ability to probe these phenomena through polarimetric studies spanning from millimeter to X-ray wavelengths.
In particular, the launch of the Imaging X-ray Polarimetry Explorer (IXPE) has enabled the first systematic measurements of X-ray polarization in blazars \citep{Weisskopf2022}. IXPE observations of HSP blazars \citep{Liodakis2022,DiGesu2023,ixpedata1} have revealed X-ray polarization degrees ranging from 10\% to 20\%, showing a systematic decrease at lower frequencies, with the optical polarization degree systematically larger than the millimiter one.
This phenomenology, and the one observed in the  RoboPol sample, has been explained by energy stratification of emission regions as electrons lose energy via radiation \citep{Liodakis2022}, providing strong support for shock acceleration models \citep{Kirk2000} and suggesting a complex multi-zone structure within blazar jets where different energy bands probe distinct physical regions. However,  IXPE observations have also revealed EVPA rotations in several blazars \citep{Middei2023}, which present a challenge to simple energy-stratified models that would predict stable, frequency-independent EVPAs if magnetic field configurations remain coherent across emission zones. 
This discrepancy suggests either that the magnetic field geometry varies significantly between different emission regions or that additional physical processes, such as Faraday rotation or turbulent magnetic field fluctuations, play important roles in determining the observed polarization characteristics.

In this work, we present a comprehensive Monte Carlo framework for modeling multi-zone synchrotron emission in blazar jets, specifically designed to reproduce the energy-dependent polarization characteristics observed from millimeter to X-ray wavelengths. 
Our approach incorporates spatially resolved emission regions with distinct physical properties, in a purely turbulent framework, that is, we do not assume any correlation between magnetic field and cell size and/or position; we only assume a power-law distribution for the cell size, and we tested different levels of correlation between the size of the cells and cutoff in the electron distribution. This minimal approach aims to provide the most likely statistical characterization of the cell distributions, in terms of size and energy of the electrons, able to reproduce the energy-dependent polarization patterns observed in the millimiter-to-X-ray data for IXPE HPS, and in the optical for the RoboPol dataset. 
The paper is organized as follows: in Sect. \ref{sec:single-cell} we discuss the single-cell S emission, and we provide some useful $\delta$ approximations both for the S SED emission and for the fraction polarization.
In Sect. \ref{sec:N-cells-no-flux-dispersion} we investigate analytical trends for the fractional polarization, for the case of a multi-zone scenario, without flux dispersions. In Sect. \ref{sec:N-cells-with-flux-dispersion}, we extend the results of Sect. \ref{sec:N-cells-no-flux-dispersion} to the case of dispersion of cells flux, and we introduce the concept of flux-weighted effective number of cells contributing to the observed polarization.
In Sect. \ref{sec:sim_workflow} we describe the setup for our Monte Carlo (MC) simulation workflow. 
In Sect. \ref{sec:sim_trends} we analyze our MC results for the case of no dispersion on cells' flux and same cells size (Sect. \ref{sec:sim-ss}), and the case of dispersions on cells' flux and power-law-distributed cells' size (Sect. \ref{sec:sim-PL}), presenting the predicted trends for the patterns of the multiwavelength fractional polarization. 
In Sect. \ref{sec:data_comparison} we compare our trends to the multiwavelength trends observed, for a sample of HSP, by IXPE, and to the optical trends observed by RoboPol for a large sample of $\gamma$-ray-loud blazars. Finally, in Sect \ref{sec:disc_concl} we discuss our results and present our conclusions.

\section{Synchrotron polarization}
\subsection{Single cell}
\label{sec:single-cell}

Let us consider a single spherical emitting region, with an ordered magnetic field $B$, and a population of relativistic electrons, described by energy distribution (EED) $n(\gamma)$.
According to the standard synchrotron (S) theory \citep{Westfold1959,Ryb1986}, the degree of linear polarization is given by the flux emitted  in the  directions parallel ($F_{\nu,\|}$) and perpendicular ($F_{\nu,\perp}$)  to the projection of the magnetic field on the plane of the sky,
\begin{equation}  
   \Pinuord=\frac{F_{\nu}^{\rm pol}}{F_{\nu}}=\frac{F_{\nu,\perp}-F_{\nu,\|}}{F_{\nu,\perp}+F_{\nu,\|}}=
   \frac{\int G(\nu/\nu_c)n(\gamma) d\gamma}{\int F(\nu/\nu_c)n(\gamma) d\gamma},
   \label{eq:pol_definition}
\end{equation}
where $\nu_c$ is the S critical frequency, and the functions  $F(x) \equiv x \int K_{5/3}(\xi) d\xi$ and $G(x)\equiv xK_{2/3}(x)$, are defined in terms of modified Bessel functions of the second kind and fractional order.
The S numerical computation, both for the SED and polarization, is performed using the \jetset \footnote{\url{https://github.com/andreatramacere/jetset}},  \footnote{\url{https://jetset.readthedocs.io/en/latest/index.html}} code \texttt{ v1.3.1} \citep{Tramacere2009, Tramacere2011, Tramacere2019}.
If we assume a reference direction on the sky plane, and $\chi$ is the angle between this direction and the projection of the magnetic field on the plane of the sky, then $\Pinuord$ can be expressed in terms of Stokes' parameters:
\begin{equation}
   \Pinuord= \frac{\sqrt{(U_{\nu}^2 + Q_{\nu}^2)}}{F_{\nu}}=
   \frac{F_{\nu}^{\rm pol}  \sqrt{\sin(2\chi)^2 + \cos(2\chi)^2}} {F_{\nu}}.
   \label{eq:stoke_definition}
\end{equation}
For a power-law EED, with a spectral index $p$,
\begin{equation}
   \Pi(p)=\frac{p+1}{p+7/3}.
   \label{eq:pol_asym}
\end{equation}
The relation above is valid only for a pure power law, or as long as the S emission is dominated by the power-law branch of the underlying EED. 
We can obtain a more generic trend, for Eq.(\ref{eq:pol_asym}), by deriving an approximate expression of $\Pi_{\nu}(p)$ using the standard S theory in $\delta-$function approximation,
\begin{equation}  
   \begin{aligned}
      \nu  &\approx& 10^6 \gamma^2 B \delta /(1+z) \\
      F_\nu &\approx&  V\gamma n(\gamma)B \delta^3(1+z)/(4 \pi d_L(z)^2)\\
      \nu F_\nu   &\approx&  V\gamma^3 n(\gamma)B^2 \delta^4/(4 \pi d_L(z)^2), 
   \end{aligned}
   \label{eq:delta_approx}
\end{equation}
where $z$ is the cosmological redshift, and $\delta$ the beaming factor, $V$ is the volume of the emitting source, and $d_L$ is the luminosity distance.
These relations link the flux emitted at a given frequency to the corresponding shape of the electron distribution at $\gamma(\nu)$.
The peak frequency of the SED,$\nu_p$, will be at
\begin{equation}
   \begin{aligned}
   \gamma_{3p} &=& {\rm peak ~~of}~~ \gamma^3 n(\gamma)\\
   \nu_p &\approx& 10^6 \gamma_{3p}^2 B \delta (1+z).  
   \end{aligned}    
   \label{eq:delta_approx_nu_p}
\end{equation}
Assuming that EED is a power-law with an exponential cutoff,
\begin{equation}
   \begin{aligned} 
   n(\gamma)&=&K \gamma^{-p}\exp(-\gamma/\gamma_{\rm cut})\\
   \gamma_{3p}&=&\gamma_{\rm cut} (3 -p), {~~~~\rm for~~~}  p<3\\
   \nu_p &\approx& 10^6 {\gamma_{\rm cut} (3 -p)}^2 B \delta (1+z).  
   \end{aligned}    
   \label{eq:delta_approx_n_gamma}
\end{equation} 
By using the $\delta-$function approximation, the energy-dependent slope, $p_{\nu}^{(\delta)}$, can be evaluated as the log-derivative of $n(\gamma)$ at $\gamma(\nu)$,
\begin{equation}
   p_{\nu}^{(\delta)} = \frac{d \log{n(\gamma)}}{d \log{\gamma}} \bigg \rvert _{\gamma(\nu)} 
   \label{eq:p-delta-approx-n}
\end{equation}
or, conversely,  $p_{\nu}^{(\delta)}$ can be estimated from $F_{\nu}$ as
\begin{equation}
   p_{\nu}^{(\delta)}= 2 \frac{d \log{F(\nu)}}{d \log{\nu}}+1.
   \label{eq:p-delta-approx-p-log-deriv}
\end{equation} 
Finally, the $\delta-$function approximation for the fractional polarization can be obtained from Eq.(\ref{eq:pol_asym}):
\begin{equation}
   \Pi_{\nu}^{\delta}=\Pi(p_{\nu}^{(\delta)}).
   \label{eq:pol-delta-approx-nu}
\end{equation}

\begin{figure}
   \centering
   \includegraphics[width=.85\columnwidth]{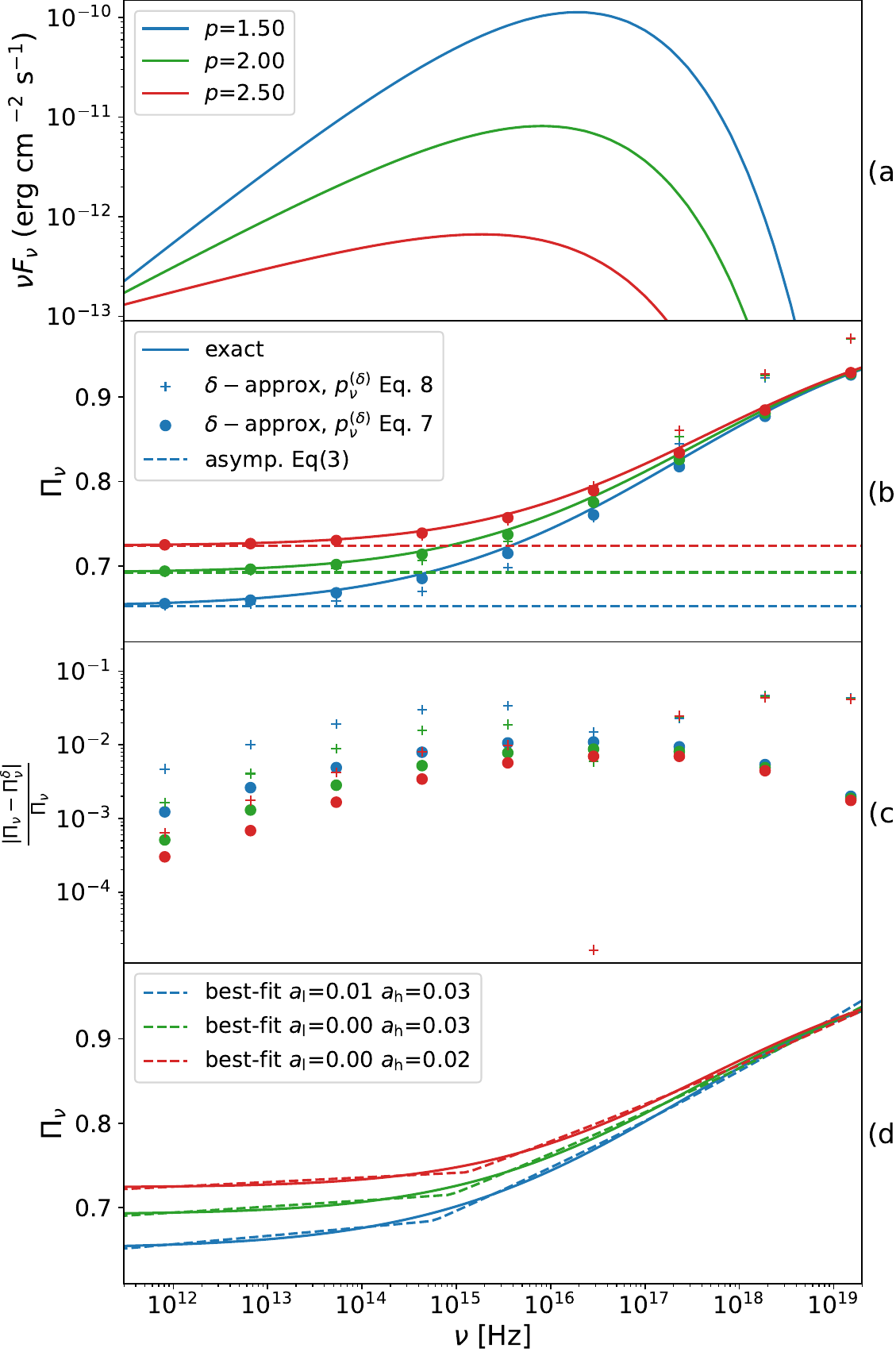}
   \caption{\emph{Left panels:} The S SED (\emph{row a}), and the corresponding fractional polarization (\emph{row b}) as a function of the frequency, for a power-law cutoff EED, $\gamma_{\rm cut}=5\times 10^4$, and $p=[1.5,2.0,2.5]$. The solid lines in the \emph{row b} panel mark the $\Pinuord$ evaluated using Eq.(\ref{eq:pol_definition}), the circles mark the $\delta$ approximation from Eq. (\ref{eq:pol-delta-approx-nu}) using $p_{\nu}^{(\delta)}$ from  Eq. (\ref{eq:p-delta-approx-n}), and the crosses using $p_{\nu}^{(\delta)}$ from  Eq. (\ref{eq:p-delta-approx-p-log-deriv}). The dashed lines mark the asymptotic power-law trend from Eq. (\ref{eq:pol_asym}). The relative error for the $\delta-$approximation methods, w.r.t. the 
   exact method (\emph{row c}). In the \emph{row d} panel, we report the best-fit of $\Pinuord$ by means of a broken power-law function.} 
   \label{fig:sing_cell_pol}
\end{figure}

In Figure \ref{fig:sing_cell_pol}, we show a comparison for $\Pinuord$  using the different methods described above. The row-a panel shows the S SED, and the row-b panel the corresponding fractional polarization as a function of the frequency, for a power-law cutoff EED, for $\gamma_{\rm cut}=5\times10^4$, $\gamma_{\rm min}=2$, and $p=[1.5,2.0,2.5]$.
We notice that the exact calculation of  $\Pinuord$, evaluated using Eq.(\ref{eq:pol_definition}), and marked by solid lines, is well approximated by the  $\delta-$approximation from Eq. (\ref{eq:pol-delta-approx-nu}),
both using $p_{\nu}^{\delta}$ from Eq. (\ref{eq:p-delta-approx-n}) (solid circles), and using $p_{\nu}^{\delta}$ from Eq. (\ref{eq:p-delta-approx-p-log-deriv}) (crosses). 
The fractional deviation of the two $\delta-$approximation methods, reported in the c-row panel of Figure \ref{fig:sing_cell_pol}, reaches a maximum value of a few percents, with the method using $p_{\nu}^{\delta}$ from Eq. (\ref{eq:p-delta-approx-p-log-deriv}) providing a better performance compared to $p_{\nu}^{\delta}$ from Eq. (\ref{eq:p-delta-approx-n}).
We also notice the asymptotic behavior of $\Pinuord$: below $\nu_p$ $\Pinuord$ asymptotically approaches the value dictated by Eq. (\ref{eq:pol_asym}) (horizontal dashed lines), whilst, above $\nu_p$ it deviates significantly from the  Eq. (\ref{eq:pol_asym}), asymptotically approaching the maximal polarization limit of 1.0, for $\nu/\nu_p>>1$.
More in detail, if we define the polarization slope, $a$, as the log-derivative:
\begin{equation}
   a=\frac{d \log{\Pinuord(\nu)}}{d \log{\nu}}
\label{eq:pol-sample}
\end{equation} 
and we fit $\Pinuord$ by means of a broken power-law, we notice that  $\Pinuord$, below $\nu_p$, asymptotically approaches a value of $a_{\rm l}\approx 0$,  above $\nu_p$ reaches a value of 
$a_{\rm h} \approx [0.02,0.03]$. 
The maximum value of $a$  is limited by the log-ratio of the maximum excursion of polarization(typically $[0.75-1]$) to the maximum SED excursion in frequency above $\nu_p$ (typically a few decades, depending on the EED).

\subsection{Synchrotron polarization: case of N  cells with no dispersion on flux}
\label{sec:N-cells-no-flux-dispersion}

Now, let's consider a system of $N_c$  cells, with  $N_\nu\leq N_c$ indicating the number of cells emitting at the frequency $\nu$. In each cell, the magnetic field is ordered, but the single-cell position angle,  $\chi_i^{\rm r}$, is randomly oriented, and $F_{\nu,i}$ is the single-cell flux. 
We assume that the pitch angle (the angle between an electron's velocity vector and the direction of the magnetic field), in each cell, it can be either fixed or distributed uniformly, without any loss of generality, since in the former case the term in $\sin(\alpha)$ can be factored out in Eq. \ref{eq:pol_definition}, and in the latter, the average of a uniform distribution of $\alpha$, does not change the results compared to the fixed angle case.
For the specific case of almost identical cells, that is, $F_{\nu,i}\approx\bar F_\nu$, and  $F_{\nu,i}^{\rm pol}\approx\bar F_{\nu}^{\rm pol}$, the average degree of the linear S polarization ($\Pinuave$), at a given frequency, can be evaluated as in \cite{marscher2022}: 
\begin{equation}
   \begin{split} 
   \Pinuave&= \Pinuord \bar F_\nu<\frac{\sqrt{ 
    (\sum_i^{N_\nu}  \sin(2\chi^r_i))^2 + (\sum_i^{N_\nu} \cos(2\chi^r_i))^2}}
   {\sum_i^{N_\nu} \bar F_{\nu}}>\approx \\ 
   &\approx \frac{\Pi^{\rm ord}_\nu}{\sqrt{ N_\nu}}
   \end{split} 
   \label{eq:frac_pol_same_cells}
\end{equation}

We can also define a polarization factor, as in \cite{marscher2022}:
\begin{equation}
\kappa_\nu =   \sqrt{f_{\rm ord}^2+\frac{(1-f_{\rm ord})^2}{N_\nu}}
\end{equation}
where $f_{\rm ord}$ is the fraction of coherent magnetic field in the system, which reduces to $\kappa_\nu = \frac{1}{\sqrt{N_\nu}}$, for $f_{\rm ord}=0$.
In the following, we investigate the case of $f_{\rm ord}=0$, hence, all the results presented in the following can be easily rescaled to the case of $0<f_{\rm ord}\leq 1$.

\subsection{Synchrotron polarization: Case of N cells with dispersion on flux}
\label{sec:N-cells-with-flux-dispersion}
In a more general and realistic scenario, the flux of each cell, $F_{\nu,i}$, depends on the physical properties of each cell, that is, $F_{\nu,i}\neq\bar F_\nu$, preventing the $F_\nu$ term from being factored out in Eq. \ref{eq:frac_pol_same_cells} :
\begin{equation}
   \Pi_\nu= \frac{\sqrt{(\sum_i^{N_\nu} F^{\rm}_{\nu,i} \Pi_{\nu,i} \sin{2\chi_i})^2 + (\sum_i^{N_\nu} F_{\nu,i}\Pi_{\nu,i}\cos{2\chi_i})^2}}{\sum_{i}^{N_\nu}F_{\nu,i}}.
   \label{eq:frac_pol_generic}
\end{equation}
This implies that the \emph{effective} number of emitting cells at a given frequency will depend on the statistical distribution of the physical parameters of the cells in the system. 
From a statistical standpoint, the \emph{effective} number of emitting cells can be  interpreted as the effective sample size (ESS) in a sample with weights ($w_i$), using the approach of \cite{samplesize}:
\begin{equation}
  {\rm ESS}= \frac{(\sum_{i}^{N_\nu} w_{\nu,i})^2}{\sum_{i}^{N_\nu} w_{\nu,i}^2},
\end{equation}
where the cell weights, for Eq. \ref{eq:frac_pol_generic}, are defined as:
\begin{equation}
   w_{\nu,i}=\frac{F_{\nu,i}}{\sum F_{\nu,i}},   
\end{equation}
hence,  the effective number of cells contributing to a given frequency, $N_{\nu}^{\rm eff}$, will read:
\begin{equation}
   N_{\nu}^{\rm eff} = \frac{(\sum_{i}^{N_\nu} w_{\nu,i})^2}{\sum_{i}^{N_\nu} w_{\nu,i}^2} \leq N_{\nu}   \rightarrow
   \kappa_\nu^{\rm eff} = \frac{1}{\sqrt{N_{\nu}^{\rm eff}}},
\label{eq:N_nu_eff}
\end{equation}
and the average frequency-dependent fraction polarization will read:
\begin{equation}
   \Pinuave \approx \frac{<\Pi^{\rm ord}_\nu>} {\sqrt{N_{\nu}^{\rm eff}}} = \kappa^{\rm eff}_\nu <\Pi^{\rm ord}_\nu>.
   \label{eq:Pi_nu_ave_eff}
\end{equation}
We stress that our definition of $N_{\nu}^{\rm eff}$ provides a more robust estimate of the flux-weighted cell contribution to the total fractional polarization compared to that presented in \cite{Peirson2019}, being the latter defined as the number of zones contributing half of the integrated flux.

The relevant result here is that the observed fractional polarization, in general, will not depend on $N_{\nu}$, but on $N_{\nu}^{\rm eff}$. As a consequence, the estimate of $N_{\nu}$ (or $N_c$), from $(\Pinuord / \Pinuave)^2$ will lead to an underestimation of the actual number of cells in the system. 

\section{Simulation setup and workflow}
\subsection{MC simulations setup}
\label{sec:sim-setup}
In a turbulent medium, even though we might have similar cell sizes, the EED, the beaming factor, and other parameters can change. We need to accurately compute the S emission and polarization for each cell and generate the individual cell properties via a Monte Carlo approach. We assume that our system has a spherical geometry, with a characteristic radius, $R_S$, and that the single-cell radius, $R_c$, is distributed as a power-law, with index $q$,  via the scaling factor $r$, according to:
\begin{equation}
\begin{aligned}
   r  &\sim& {r}^{q},&~~~ q\leq 0, ~~~0<r_{\rm min}<r<r_{\rm max}\leq 1\\
   f_{R_c}&=& r^{q}R_{S}& 
   \label{eq:R_pdf}
\end{aligned}
\end{equation} 
we set $R_S=10^{16}$ cm.
The cell values of the magnetic field intensity, $B_c$, and of the beaming factor, $\delta_c$, are distributed as a flat PDF or  have a fixed value:
\begin{equation}
 \begin{aligned}
   f_{\delta_c} &=& \mathcal{U}[\rm \delta_{\rm min},\delta_{\rm max}], &\mathrm{~or}& f_{\delta_c} =\delta_0\\
   f_{B_c} &=& \mathcal{U}[B_{\rm min},B_{\rm max}], &\mathrm{~or}& f_{B_c} =B_0\\
 \end{aligned}
\end{equation}
The  position angle of the magnetic field, in a single cell, $\chi^{\rm r}_c$, is randomly oriented: 
$f_{\chi^{\rm r}_c} = \mathcal{U}[0,2\pi]$.

The EED is assumed to be  a power-law with an exponential cutoff:
\begin{equation}  
   n(\gamma)=K \gamma^{-p}\exp(-\gamma/\gamma_{\rm cut}),
   \label{eq:n_gamma}
\end{equation} 
the cell values of the index EED index, $p_c$, are distributed as a uniform distribution, or  have a fixed value:
\begin{equation}
    f_{p_c} = \mathcal{U}[p_0,p_0+\Delta p], \mathrm{~or~} f_{p_c} =p_0.
\end{equation}
\begin{figure*}
   \centering
   \includegraphics[width=1\textwidth]{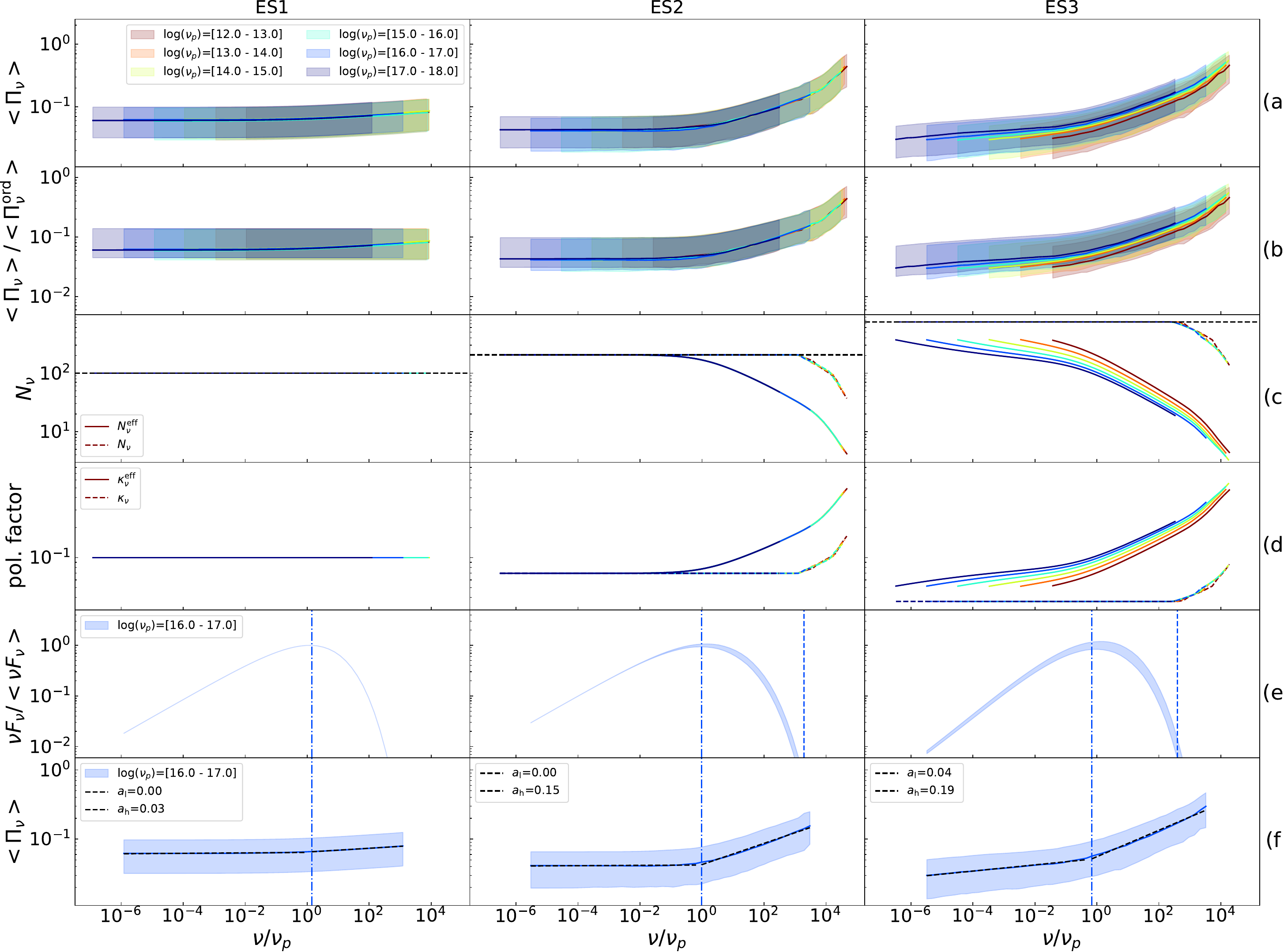}\\
    
   \caption{Different colors identifying the corresponding $\nu^p$ bins, which are the same for all the panels.
      \emph{Top panels:} \emph{Left column:} case of identical cells (configuration ES1 in Table \ref{tab:sim_conf_ss}). 
   \emph{Middle column:} case of configuration ES2 in Table \ref{tab:sim_conf_ss}, i.e. using 
       $\fgammacut$=\emph{log-uniform}. \emph{Right column:} panel, case of configuration ES3,
   same as ES2 but adding a flat PDF, for $p:f_{p}\sim\mathcal{U}[1.8,2.8]$.
   All the trends are reported versus $\nu/\nu_p$. Solid lines mark the MC 0.5 quantiles and shaded areas mark the $1$-$\sigma$ quantiles dispersion, for the MC trials. 
   \emph{Row a}: the trend of the ratio of $\Pinuave$.
   \emph{Row b}: same as for Row 1, but for trials-averaged trends for   $\Pinuave/\Pinuaveord$.  
   \emph{Row c}: trials-averaged trends for  $N_{\nu}$ (dashed lines) and $N_{\nu}^{\rm eff}$ (solid lines).  
   \emph{Row d}: trials-averaged trends for  $\kappa_\nu$ (dashed line) and $\kappa_\nu^{\rm eff}$ (solid lines).
   \emph{Row e}: SEDs for the $\nu_p$ bin =[$10^{16}-10^{17}$] Hz, the dashed vertical line shows the  $\nu/\nu_p$ value, above which not all the cells contribute to the SED flux. The dot-dashed vertical lines mark  
   $\nu=\nu_p^{\rm min}$, where $\nu_p^{\rm min}$ is the SED peak value for the lowest flux cell.}
   \label{fig:pol_ss_trends}
\end{figure*}
We investigated four different scenarios for the PDF of $\gamma_{\rm cut}$, $\fgammacut$:
a scenario with a fixed value of $\gamma_{\rm cut}$:
\begin{equation}
   \emph{$\delta_f$}:\fgammacut={\gamma}_{{\rm cut}}^{\rm ref}\\
\end{equation}
and a scenario with dispersion on $\gamma_{\rm cut}$.
For the latter scenario, two distributions have no correlation between cell size and $\gamma_{\rm cut}$:
\begin{equation}
\begin{aligned}
   \emph{uniform}&:&\fgammacut&=\mathcal{U}[\gamma_{\rm cut}^{\rm LB},{\gamma_{{\rm cut}}^{\rm ref}}]\\
   \emph{log-uniform}&:&\fgammacut&=\mathcal{U}[\log(\gamma_{\rm cut}^{\rm LB}),\log({\gamma_{{\rm cut}}^{\rm ref}})] \\
   &~&~&=\frac{1}{\gamma_{\rm cut}\log({\gamma_{{\rm cut}}^{\rm ref}}/\gamma_{\rm cut}^{\rm LB})}\propto \gamma_{\rm cut}^{-1}\\
\end{aligned}
\end{equation}
and the third one  establishes a linear relationship between $R_c$ and $\gamma_{\rm cut}$, 
setting $ \gcutminratio=r_{\rm min}$ and $\gamma_{{\rm cut}}^{\rm ref}=r_{\rm max}$, and applying the transformation $\gamma_{{\rm cut}}\propto\gamma_{{\rm cut}}^{\rm ref}r$, leading to:
\begin{equation}
\emph{linear}:     
\fgammacut \propto \frac{\gamma_{\rm cut}^{q}}{(\gamma_{{\rm cut}}^{\rm ref})^{q+1}}
\end{equation}

The dispersion on $\gamma_{\rm cut}$ is controlled by the value of the lower bound defined as $\gamma_{\rm cut}^{\rm LB}$, as $\max(100, {\gamma_{{\rm cut}}^{\rm ref}} \gcutminratio)$, implying that lower values of  $\gcutminratio$ lead to a larger dispersion.
The \emph{linear} distribution is motivated by results from magnetic reconnection  Particle-in-Cell (PIC) simulations \citep{Limagrec, Sironi2016}, finding the EED high-energy cutoff scaling linearly with the plasmoids' width.
We notice that for $q<0$  both the \emph{log-uniform} and \emph{linear} cases result in $\fgammacut$ being a decreasing function of $\gamma_{\rm cut}$, with the relevant difference that the \emph{linear} case introduces a positive correlation between the cell size and the value of $\fgammacut$.

\subsection{MC simulations workflow}
\label{sec:sim_workflow}
For each combination of the input parameters described in Sect. \ref{sec:sim-setup}, we first perform a calibration stage, that is, we calibrate $N_c$ and $\gamma_{\rm cut}^{\rm ref}$, to obtain a trial-averaged optical polarization fraction, $\Pinuaveopt$, within $[0.4-0.5]\%$, and a trial-averaged value of $\nu_p^S$ within=$[0.8-1.2] \times 10^{17}$ Hz. 
As a reference value for the optical frequency, we use the value of $5\times 10^{14}$ Hz. Details of the calibration stage are reported in Sect. \ref{sec:calibration_stage}.
At the end of the calibration stage, the calibrated values of $N_c$ and $\gamma_{\rm cut}^{\rm calib}$ are used for the full MC simulations as follows:
\begin{itemize}
   \item  we define six bins for $\nu_p^S$, evenly spaced in the logarithm, over the range [$\nu^S_{p,\rm min}=10^{12}$, $\nu^S_{p,\rm max}=10^{18}$] Hz. Each value of $\nu_{p,i}^S$ is mapped to a corresponding bin for the EED cutoff energy of ${\gamma}_{{\rm cut},i}^{\rm ref}$, by scaling the calibrated value of $\gamma_{\rm cut}^{\rm calib}$ according to the third of Eq. \ref{eq:delta_approx_n_gamma}.

   \item  For each bin of ${\gamma}_{{\rm cut},i}$ we run 1000 trials using always the same value of $N_c$, 
   and drawing ${\gamma}_{{\rm cut}}^{\rm ref}$ from $\mathcal{U}[\gamma_{{\rm cut},i}^{\rm ref},\gamma_{{\rm cut, i+1}}^{\rm ref}]\\$
   
   \item For each run, we verify that the sample variance converges according to Stein’s two-stage scheme 
   \citep{stein1945} as implemented in \cite{mc_size}. In detail, we have verified, for each run, that the interval, $[\Pinuave (1 - \varepsilon_{\rm MC}), \Pinuave (1 + \varepsilon_{\rm MC})]$, for $\varepsilon_{\rm MC} \leq 0.15$, provides confidence interval for the true value of $\Pi_{\nu}$, at a confidence level of $0.95$, up to $\nu\leq 1000\times \nu_p$.
\end{itemize}
At the end of the MC, we have 1000 realizations for each of the six $\nu_p^S$ bins, where the only changed parameter per bin is $\gamma_{\rm cut}^{\rm ref}$, that is,  each trial and each bin of $\nu_p^S$, has the same value of $N_c$ and the same PDFs $f_{R_c}, f_{\delta_c}, f_{B_c}$, and $f_{\chi^r}$.
The trials are eventually binned according to the $\nupSave$ value in the $\nu_{p,i}^S$ bins, and the relevant statistics of the parameters are extracted.

\begin{table}[]
   \caption{MC parameter space for  equal-size distributed cells.}
         
   \begin{tabular}{l|c|c}
   \hline
   configuration         &$\fgammacut$    & $f_{p_c}$                       \\ \hline
   ES1                   &\emph{$\delta_f$}         & 2.3                       \\
   ES2                   &\emph{log-uniform}        & 2.3                       \\
   ES3                   &\emph{log-uniform}        & $\mathcal{U}[1.8,2.8]$    \\
   \end{tabular}
   \tablefoot{For all the configurations, we use:  $q=0$, $r_{\rm min}=r_{\rm max}=0.1$,  $f_{B_c} =B_0=0.1$, $f_{\delta_c} =\delta_0=10$, $f_{\chi^{\rm r}_c}=\mathcal{U}[0,2\pi]$,
   $\gcutminratio=0.1$, and we perform 1000 trials.}
   
\label{tab:sim_conf_ss} 
\end{table}

\begin{figure*}
   \centering
   \includegraphics[width=\textwidth]{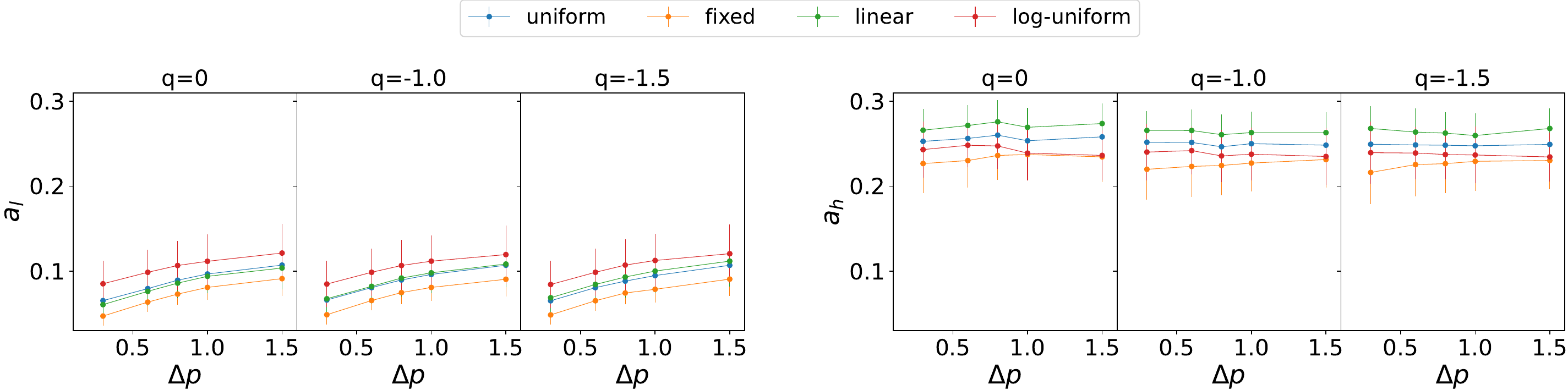}\\
   \vspace{.2cm}
   \includegraphics[width=\textwidth]{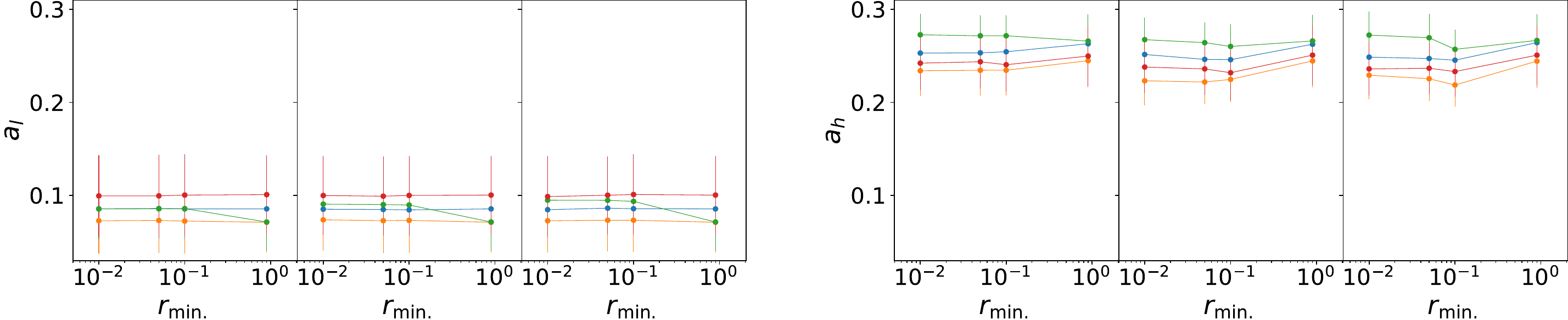}\\
   \vspace{.2cm}
   \includegraphics[width=\textwidth]{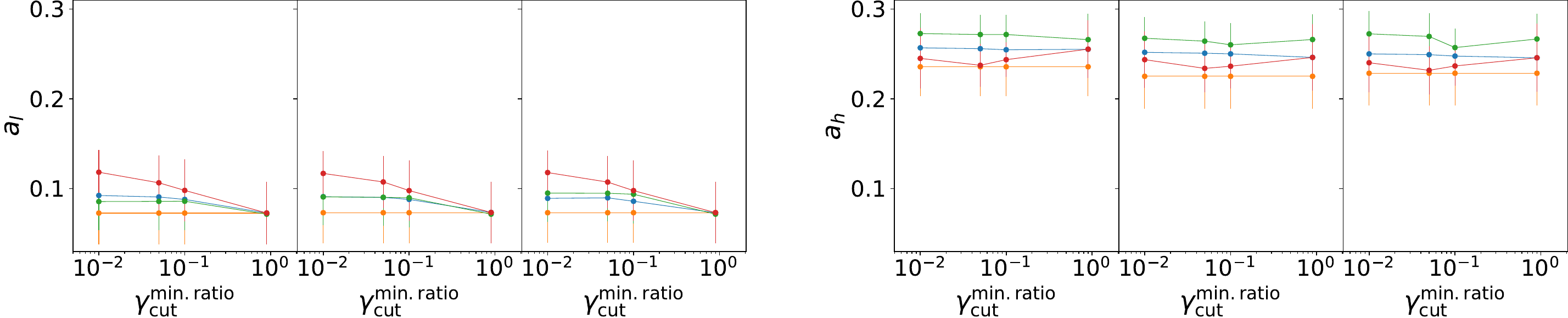}\\
   \caption{
   Summary of the polarization slope trends for the PL-distributed parameter space reported in Table \ref{tab:sim_conf_pl}. Each subpanel refers to the three different values of $q=[0,-1.0,-1.5]$, as reported in the subpanel title. \emph{Top panels:} Low-energy ($a_l$, left panels) and high-energy ($a_h$, right panels) polarization slopes as a function of the dispersion in the electron energy distribution index, $\Delta_p$. \emph{Middle panels:} $a_l$ and $a_h$ as a function of the minimum cell size ratio, $r_{\rm min}$. \emph{Bottom panels:} $a_l$ and $a_h$ as a function of the minimum cutoff ratio, $\gcutminratio$. Different colors correspond to the different $\fgammacut$ PDF reported in the top legend.}
   \label{fig:pol_full_mc}
\end{figure*}
\section{Simulation trends}
\label{sec:sim_trends}
In this section, we provide a statistical description of the $\Pinuave$ trends, for two different scenarios: 
the case of cells with the equal size (Sect. \ref{sec:sim-ss}) and the case of power-law-distributed cell sizes
(Sect. \ref{sec:sim-PL}). We aim to do the following: 
\begin{itemize}
   \item verify the  $\Pinuave$ trend provide by Eq.\ref{eq:Pi_nu_ave_eff} and in particular the connection
   between the polarization factor $\kappa^{\rm eff}_\nu $ and $N_{\nu}^{\rm eff}$.
   \item characterize the  $\Pinuave$ trends as a function of $\nu/\nu_p$
   \item identify how the different PDFs for  $f_{R_c}, f_{\delta_c}, f_{B_c}$, and $\fgammacut$ impact the trends.
  \end{itemize}
\subsection{Case of equal-size cells}
\label{sec:sim-ss}
We start by investigating the case of equal-size (ES) cells with three different configurations, whose parameters are summarized in Table \ref{tab:sim_conf_ss}.
In Figure \ref{fig:pol_ss_trends} we show the impact, on the frequency-dependent polarization pattern $\Pinuave$, of different levels of randomization, from configuration ES1 to ES3.
The left column panels refer to the case of identical cells, i.e., the configuration ES1 of Table \ref{tab:sim_conf_ss}. 
The middle column panels refer to the configuration ES2, i.e. same as ES1 but using for  $\fgammacut$ the \emph{log-uniform} PDF and, finally, the right column panels refer to the ES3 case, with the same configuration as for the ES2 case, but adding a uniform PDF, for $p$, $f_{p}\sim\mathcal{U}[1.8,2.8]$.  
For all the cases, we use $\gcutminratio=0.1$.
\begin{figure*}
   \centering
   \includegraphics[width=0.8\columnwidth]{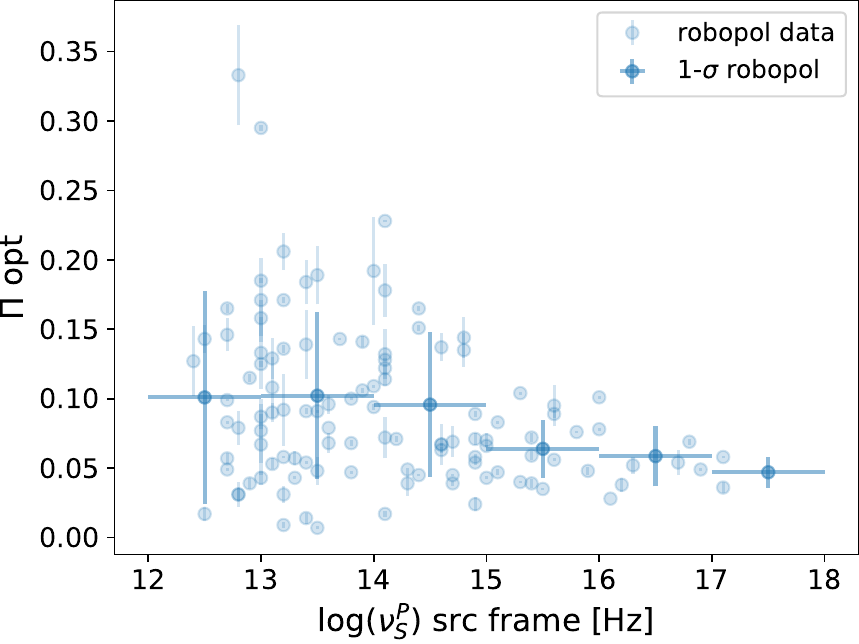}
   \includegraphics[width=0.8\columnwidth]{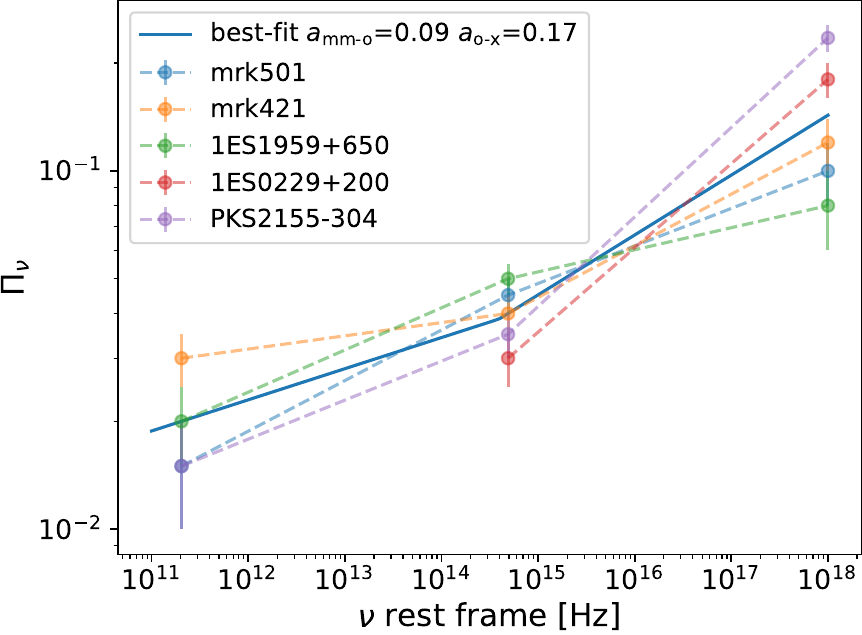}
   
   \caption{\emph{Left panel}: The limiting envelope, found by RoboPol \citep{Angelakis2016},  for a large sample of $\gamma$-ray-loud blazars between $\nu_p^S$ and both the average fractional optical polarization and its dispersion. \emph{Right panel}: The $\Pi_\nu$ trend observed in by IXPE for the HPS Mrk 501, Mrk 421, PKS2155-304, 1ES0229+200, and 1ES1959+650 collected from recent 
   IXPE multiwavelength campaigns \citep{Liodakis2022,DiGesu2023,ixpedata1,Middei2023}.}
   \label{fig:obs_data_sets}
\end{figure*}
The trends for the ratio of $\Pinuave$ and $\Pinuave/\Pinuaveord$  are shown in the panels of row \emph{a} and \emph{b}, respectively, where the shaded area represents the $1$-$\sigma$ quantiles dispersion for the MC trials, and the solid line represents the median value. 
The different colors identify the corresponding $\nu^p$ bins, and are the same for all the panels. The \emph{c}-row panels show the trends for the trials-averaged value of $N_{\nu}^{\rm eff}$, and the  \emph{d}-row panels the trials-averaged trends for the polarization factor $ \kappa_\nu$ (dashed line) and $\kappa_\nu^{\rm eff}$ (solid lines).
The SEDs for the case $\nu_p$ bin =[$10^{16}-10^{17}$] Hz are represented in the  \emph{e}-row panels, where the shaded area encompasses the entire MC range. The dot-dashed vertical lines mark the SED peak value for the lowest flux cell,$\nu=\nu_p^{\rm min}$, and the dashed vertical lines the $\nu/\nu_p$ value above which not all the cells contribute to the SED flux. 
It is clear that for the case of identical cells (ES1), $N_{\nu}$ and $N_{\nu}^{\rm eff}$ coincide, and this translates to  $\kappa_\nu = \kappa_\nu^{\rm eff}$.  
In contrast, by adding a randomization on $\gamma_{\rm cut}$ (middle column,ES2), we start to introduce a modulation on $N_{\nu}^{\rm eff}$, with  $N_{\nu}^{\rm eff}<N_{\nu}$ for $\nu\geq\nu_p^{\rm min}$, leading to a difference between the polarization factor, with $\kappa_\nu^{\rm eff}>\kappa_\nu$, for $\nu\geq\nu_p^{\rm min}$. 
This modulation is not only related to the fact that above a given $\gamma_{\rm cut}$ only a few cells are contributing, as demonstrated by the dashed lines trends related to $\kappa_\nu$, but mostly to the impact of the different cell flux contribution, which is shown by the $\kappa_\nu^{\rm eff}$ trend.
This modulation introduces a turnover in $N_{\nu}^{\rm eff}$ and  consequently in $\kappa_\nu^{\rm eff}$ and  $\Pinuave$,  located around $\nu\gtrsim\nu_p^{\rm min}$. 
The turnover still persists when we introduce a randomization on $p$ (left column, ES3).  
We model the $\Pinuave$ trend by means of a broken power-law function:
\begin{equation}
\Pinuave \propto
\begin{cases}
 {\nu}^{a_l}, & \nu < \nu_t, \\ 
 {\nu}^{a_h}, & \nu \geq \nu_t, \\ 
\end{cases}
\label{eq:bkn_pinu}
\end{equation}
with $a_l$ and $a_h$, indicating the low and high-energy indices, respectively, and $\nu_t$ the
turnover frequency, capturing the $\nu_p^{\rm min}$ position. In the following, we analyze the trends of $a_l$ and $a_h$, focusing on the $\nu_p$ bin =[$10^{16}-10^{17}$] Hz (rows \emph{e} and \emph{f} of Figure \ref{fig:pol_ss_trends}), and we find:
\begin{table}[h!]
 \caption{MC parameter space for power-law-distributed cell sizes.}
\centering
\renewcommand{\arraystretch}{1.1}
\begin{tabular}{c|c|c}
\hline
 PDF &   Parameters  &  Values \\
\hline
$f_{R_c}$: PL & $q$            & $[0, -1.0, -1.5]$ \\
              &  $r_{\rm min}$ & $[0.01, 0.05, 0.1, 0.9]$ \\
              &  $r_{\rm max}$ & $1.0$ \\
\hline
$f_{p_c}$: $\mathcal{U}[p_0,p_0+\Delta p]$  & $p_0$  &  1.5 \\
       
           & $\Delta_p$  &  [0.3,0.6,0.8,1.0,1.3,1.5]\\
\hline
$\fgammacut$: 
\(
\left\{
\begin{array}{@{}l@{}}
\text{$\delta_f$} \\
\text{\emph{uniform}} \\
\text{\emph{log-uniform}} \\
\text{\emph{linear}} 
\end{array} 
\right.
\) &   $\gcutminratio$ & $[0.01, 0.05, 0.1, 0.9]$ \\
\hline
$f_B$: $\mathcal{U}[0.1,1.0]$ &   &  \\
\hline
$f_\delta$: $\mathcal{U}[10,30]$  &   &  \\
\hline
$f_{\chi^{\rm r}_c}$: $\mathcal{U}[0,2\pi]$  &   &  \\

\end{tabular}
\tablefoot{The MC parameter space used for power-law-distributed cell sizes. The total volume of the parameter space counts 780 different configurations given by the combination of the different PDFs and different parameters listed in the table. }
\label{tab:sim_conf_pl}
\end{table}

\begin{itemize}
   \item For the ES1 configuration, since $N_{\nu}^{\rm eff} \approx N_\nu\approx N_c$, the value of the two power-law indices, $a_{\rm l}\approx 0$ and $a_{\rm h} \approx 0.03$ reflect the single-cell polarization trend, as reported in the bottom panel of Figure \ref{fig:sing_cell_pol}, i.e. asymptotically equal to the value from equation \ref{eq:pol_asym}, for $\nu<\nu_p^{\rm min}$ ($\gamma<\gamma_{3p})$ and mildly increasing for $\nu>\nu_p^{\rm min}$ ($\gamma>\gamma_{3p})$.

   \item For the ES2 configuration, the trend below $\nu_p^{\rm min}$ is the same as for ES1, since there is no modulation on $N_{\nu}^{\rm eff}$ below $\nu_p$, on the contrary, above $\nu_p^{\rm min}$ the modulation on $N_{\nu}^{\rm eff},$ introduced by the dispersion on $\gamma_{\rm cut}$, leads to a significant hardening of $a_{\rm h} \approx 0.15$.
   We notice that such a large value of $a_h$ is incompatible with the low value obtained for the single-cell configuration ($a_h \lesssim 0.03$), which hints that a significant dispersion is required in $\gamma_{\rm cut}$ (and consequently on $\gamma_{3p}$ and $\nu_p$),  to reproduce it.

   \item For the ES3 configuration, the trends above $\nu_p^{\rm min}$ are similar to the case of ES2, with $a_{\rm h} \approx 0.17$, on the contrary, below $\nu_p^{\rm min}$, the dispersion on $p$ introduces a low-energy slope $a_{\rm l} \approx 0.05$, which is significantly larger than the value of $a_{\rm l} \approx 0.01$ observed for the case of no dispersion on $p$ (ES1 and ES2). 
   The value of the slope $a_{\rm l}$ relates to the dispersion of $p$, and will be investigated in more detail in the next subsection.

\end{itemize}

\begin{figure*}
   \centering
   \includegraphics[width=.9\textwidth]{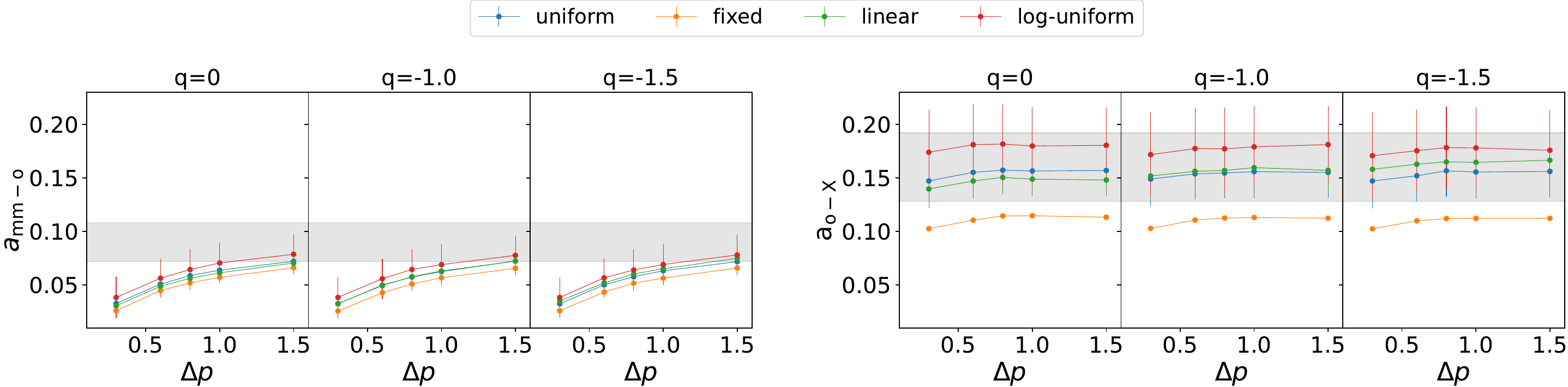}\\
   \vspace{.2cm}
      \includegraphics[width=.9\textwidth]{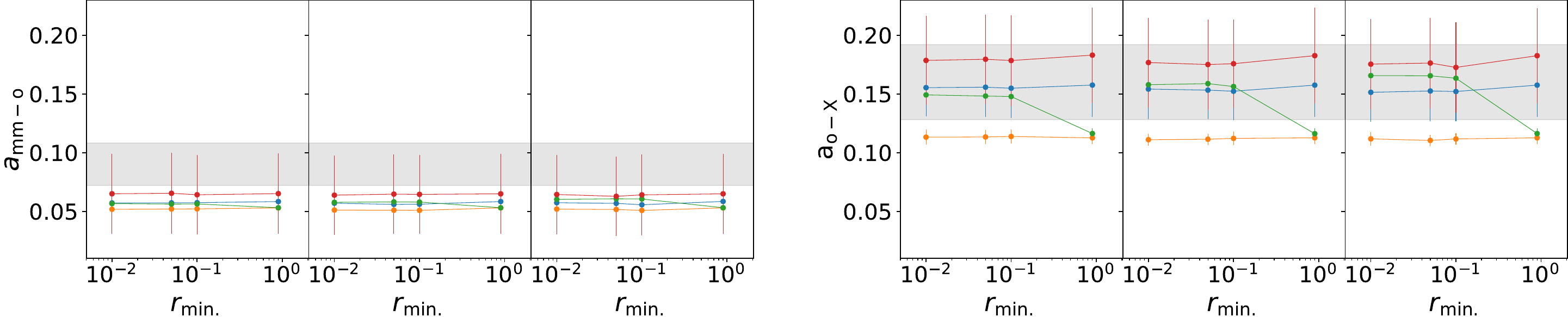}\\
   \vspace{.2cm}
   \includegraphics[width=.9\textwidth]{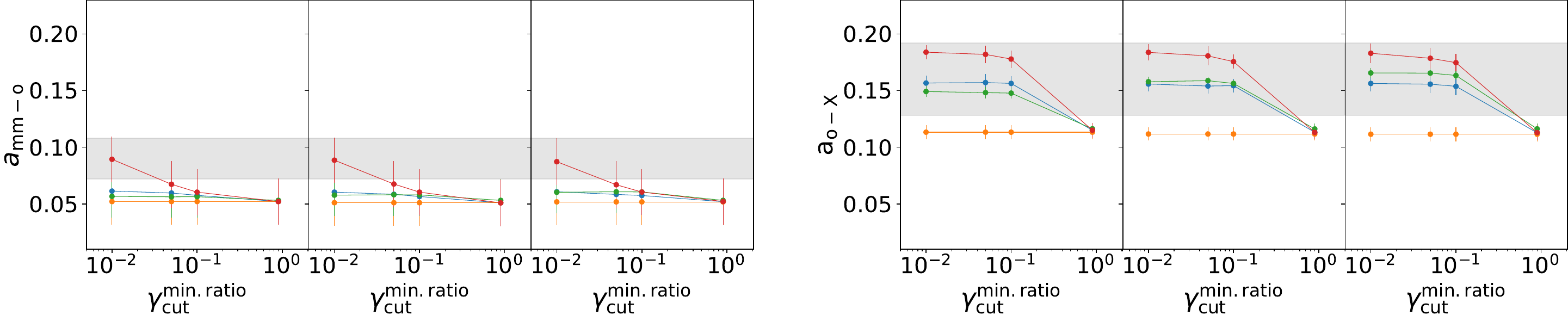}\\
   \caption{Summary of polarization slope trends for the PL-distributed parameter space reported in Table \ref{tab:sim_conf_pl}, for the HSP objects, selecting MC runs with $(10\times{15} {~\rm Hz}\leq \nu_p^S\leq 10\times{18} {~\rm Hz}$). \emph{Top panels:} Low-energy ($a_{\rm mm-o}$, left panels) and high-energy ($a_{\rm o-X}$, right panels) polarization slopes as a function of the dispersion in the electron energy distribution index, $\Delta_p$. \emph{Middle panels:} $a_l$ and $a_h$ as a function of the minimum cell size ratio, $r_{\rm min}$. \emph{Bottom panels:} $a_l$ and $a_h$ as a function of the minimum cutoff ratio, $\gcutminratio$. Different colors and line styles correspond to the various PDFs parameter configurations as indicated in the legend and in the titles of the panels.}
   \label{fig:pol_full_mc_vs_data}
\end{figure*}
In conclusion, the results of the ES scenario indicate that the slope of the high-frequency polarization is mainly influenced by the modulation of $N_{\nu}^{\rm eff}$  above $\nu_p^{\rm min}$, driven by the dispersion in $\gamma_{\rm cut}$. 
On the other hand, the low-frequency slope $a_l$ is significantly affected by the dispersion on $p$.
This behavior highlights the interplay between the EED parameters and the resulting polarization patterns, emphasizing the importance of accurately modeling the underlying physical properties of the emitting cells to interpret observational data effectively.

\subsection{Case of power-law-distributed cell sizes}
\label{sec:sim-PL}
In this section, we extend the parameter space investigated in the previous section by considering the case of cells distributed according to a power law, 
testing different $\fgammacut$ distributions and different levels of dispersion on $f_{p_c}$.
In total, the power-law (PL) scenario counts 780 different configurations, obtained by combining the different PDF configurations listed in Table\ref{tab:sim_conf_pl}.
Our goal is to investigate the impact of $f_{R_c}$, $\fgammacut$, and $f_{p_c}$ on the values of $a_l$ and $a_h$. We fit the 0.5 quantiles of the $\Pinuave$  obtained for each run using Eq. \ref{eq:bkn_pinu} and leaving all the parameters free.
In Figure \ref{fig:pol_full_mc}, we show the resulting trends of $a_l$ (left panels) and $a_h$ (right panels),  as functions of $\Delta_p$ (top panels), $r_{\rm min}$ (middle panels), and $\gcutminratio$ (bottom panels). 
Our findings on the values of $a_h$ and $a_l$ are summarized as follows:

\begin{itemize}
   \item $a_l$: Consistent with the ES configuration, we find that the average value of $a_l$ increases with the dispersion in $p$, and is largely independent of $r_{\rm min}$ and $\gcutminratio$.
   However, for the case of \emph{log-uniform} $\fgammacut$, $a_l$ decreases as $\gcutminratio$ increases. 
   This is expected since $\fgammacut\propto \gamma_{\rm cut}^{-1}$ and the cell size is uncorrelated with $\gamma_{\rm cut}$; thus, lower values of $\gcutminratio$ lead to a larger dispersion in cell flux weights at low frequencies, increasing $N_{\nu}^{\rm eff}$, with a consequent hardening of the slope of the polarization factor below $\nu_p$. 
   In contrast, for the \emph{linear} case, the positive correlation between $r$ and $\gamma_{\rm cut}$ compensates for the decreasing probability of cells with large $\gamma_{\rm cut}$ with the larger size of the same cells, mitigating the flux weight dispersion at low frequencies, and making the $a_l$ slope of the polarization factor insensitive to $\gcutminratio$.

   \item $a_h$: The high-energy slope of the polarization factor, unlike $a_l$, is independent of $\Delta_p$, and generally also of $r_{\rm min}$, since these parameters do not affect the modulation of $N_{\nu}^{\rm eff}$ above the turnover frequency. Also  $\gcutminratio$ does not impact the trends, since its impact on $\nu_p^{\rm min}$ is 
   captured by the $\nu_t$ free parameter.

\end{itemize}
Finally, in the left panel of \ref{fig:N_eff_vs_nu} we show the trend of the effective flux-averaged, $N_{\rm \nu}^{\rm eff}$, at the reference frequencies used for the  millimiter ($2\times 10^{11}$  Hz), optical ($5\times 10^{14}$ Hz),  and X-ray ($1\times 10^{18}$ Hz) frequencies, and  in the top panel of Figure \ref{fig:N_c_to_N_eff}, we show how the number of $N_c$ exceeds, at least by a factor of 10,  the value of $N_{\nu}^{\rm eff}$. The upper value of $N_c/N_{\nu}^{\rm eff}$ reaches up to a few thousand, for the X-ray  frequency, and decrease to a few hundred for the optical frequency, and up to a few tens for millimiter frequency. The effect is larger for steeper values of $q$.
 
In summary, the low-frequency polarization slope $a_l$ is most sensitive to the dispersion in $p$, while the high-frequency slope $a_h$ is mainly set by the decrease in the effective number of emitting cells above the SED peak, driven by the dispersion ob $\gamma_{\rm cut}$, and is less sensitive to the parameters related to the low-energy branch of the EED. 

\section{Comparison with observed data}
\label{sec:data_comparison}
In this section, we compare the results of our simulations for the case of PL-distributed cell size, discussed in the previous section, with the observed data. We test the following features,  observed in the phenomenology:
\begin{itemize}
   \item The $\Pi_\nu$ trend observed by IXPE for HPS objects, resulting in the X-ray polarization degrees on the order of 10\% to 20\%,  with a systematic decrease at lower frequencies, with the optical polarization degree systematically larger than the millimiter one.
   This trend is shown in the right panel of Figure \ref{fig:obs_data_sets}. The sample refers to the millimiter-to-X-ray $\Pi_\nu$ trends (dashed lines) for Mrk 421, Mrk 501, PKS2155-304, 1ES0229+200, and 1ES1959+650, collected from recent IXPE multiwavelength campaigns \citep{Liodakis2022,DiGesu2023,ixpedata1,Middei2023}. 
   The solid blue line represents a broken-power-law fit to the data, returning a millimiter-to-optical polarization index of $a_{\rm mm-o} \approx 0.1$, and an optical-to-X-ray index  of $a_{\rm o-X} \approx 0.17$
   
   \item The limiting envelope, found by RoboPol \citep{Angelakis2016}  for a large sample of $\gamma$-ray-loud blazars, between $\nu_p^S$ and both the average fractional optical polarization and its dispersion (left panel of \ref{fig:obs_data_sets}), and between  $\nu_p^S$ and the EVPA.
\end{itemize}

\subsection{IXPE HPS $\Pinuave$ trends}
To compare the HSP $\Pi_\nu$ trends observed by IXPE with our simulations, we perform the same analysis as in Sect \ref{sec:sim-PL}, but we select only the MC runs for  $10\times{15} {~\rm Hz}\leq \nu_p^S\leq 10\times{18} {~\rm Hz}$, and in place of estimating $a_l$ and $a_h$ using the best-fit values returned by Eq. \ref{eq:bkn_pinu} with all the parameters free, we compute directly the indices $a_{\rm millimiter-o} $ and $a_{\rm o-X}$ using the reference values for the millimiter, optical, and X-ray frequencies, of  $2\times 10^{11}$  Hz, $5\times 10^{14}$ Hz, and $1\times 10^{18}$ Hz, respectively.

The shaded areas in Figure \ref{fig:pol_full_mc_vs_data}, represent the $\pm 20\%$ boundary for the observed values of $a_{\rm mm-o}$ and of $a_{\rm o-X}$, and the solid lines represent the results of our MC simulations, with the error bar indicating the $2$-$\sigma$ confidence quantiles.

Although the $a_{\rm mm-o}$ index samples $\Pinuave$ below the optical frequency, which is typically below the peak frequency of the HSP objects, the pattern is similar to the one obtained for $a_l$, since the modulation in $N_{\nu}^{\rm eff}$ does not change below $\nu_p^S$. Consistent with $a_l$, we find that the most relevant parameters for the $a_{\rm mm-o}$ index are the dispersion on $p$, which is positively correlated with $a_{\rm mm-o}$, and the value of $\gcutminratio$, which has a negative correlation, for $\fgammacut$=\emph{log-uniform}. 
In general, the observed data favor a dispersion $\Delta_{p}\gtrapprox 1$, and values of $\gcutminratio\lessapprox 0.1$ for the case of $\fgammacut$=\emph{log-uniform}. We also notice that the \emph{log-uniform} distribution provides the best match with the data.

The $a_{\rm o-X}$ index shows a different trend compared to $a_h$, for $r_{\rm min}\approx 1$, for the case of \emph{linear} $\fgammacut$, and for values of $\gcutminratio\gtrapprox .1$, for all the $\fgammacut$ distributions.
The different behavior between $a_h$ and $a_{\rm o-X}$ comes from the fact that $a_h$ depends on the modulation of $N_{\nu}^{\rm eff}$ above $\nu_p^{\rm min}$. For the $a_l$ case, the modulation on $\nu_p^{\rm min}$, driven by the dispersion on $\gamma_{\rm cut}$, is captured by the $\nu_t$ free parameter. In contrast, in the case of $a_{\rm o-X}$, the value of $\nu_t$ is fixed to the value of $5\times 10^{14}$ Hz. As a consequence, the modulation on $\nu_p^{\rm min}$, driven by $\gcutminratio$, places the optical frequency below the actual value of $\nu_t$,  leading to a softening of $a_{\rm o-X}$, compared to $a_h$. This effect increases for lower values of $\gcutminratio$.
This results in a significant tension with the data for $r_{\rm min}\approx 1$, for the case of \emph{linear} $\fgammacut$, and for values of $\gcutminratio\approx 1$, for all the   $\fgammacut$ distributions, improving the constraining power of $a_{\rm o-X}$.

\begin{figure}
\centering
\includegraphics[width=.9\columnwidth]{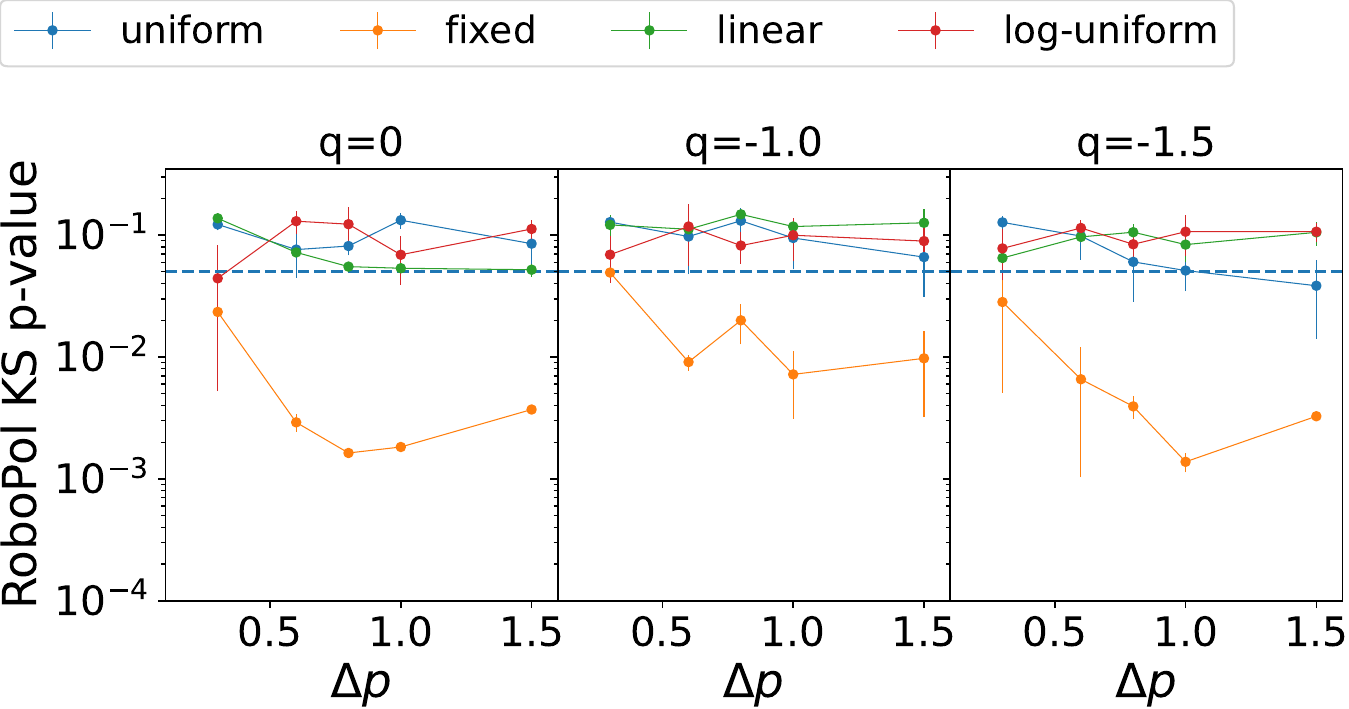}\\
\vspace{.35cm}
\includegraphics[width=.9\columnwidth]{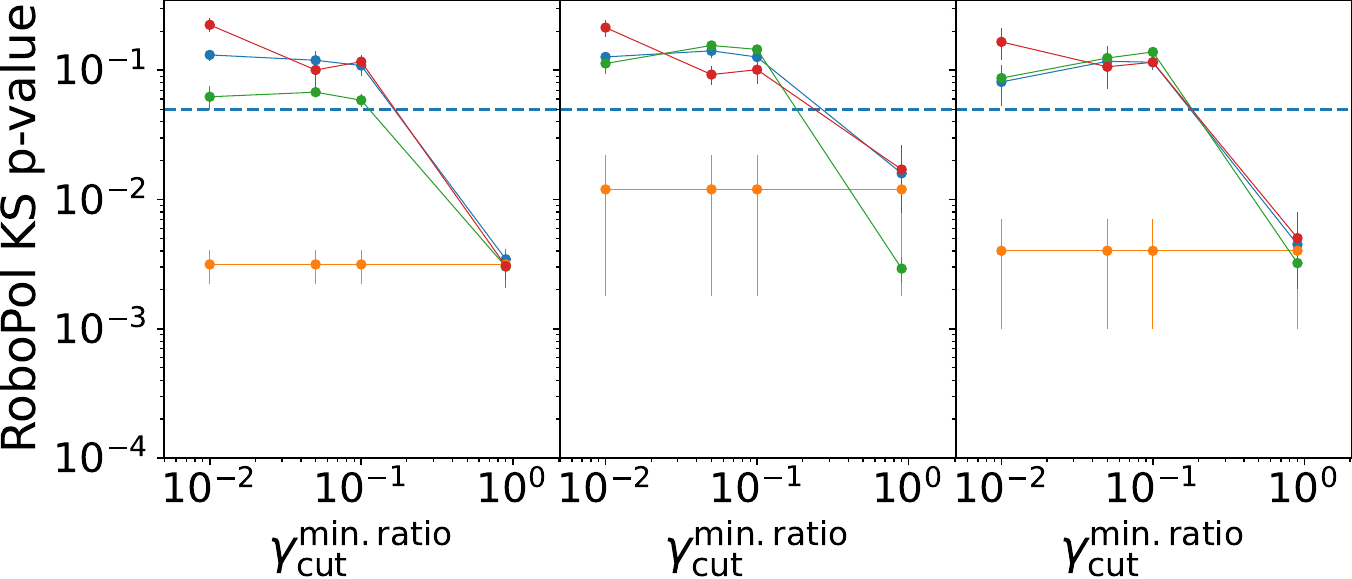}\\
\vspace{.35cm}
\includegraphics[width=.9\columnwidth]{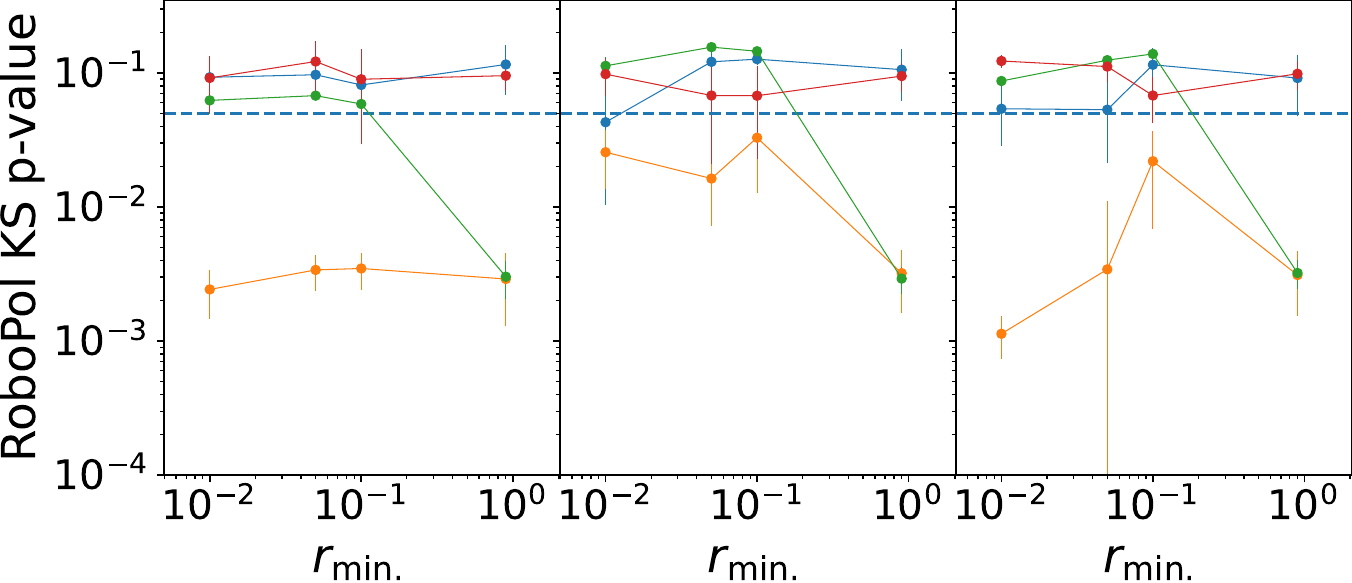}\\
\caption{Trends of the  KS test $p$ value for the $\Pi_{\rm opt}$-versus-$\nu_p^S$ limiting envelope, for the PL-distributed parameter space reported in Table \ref{tab:sim_conf_pl}. We report the KS trends  as a function of $\Delta_p$ (top panels), $r_{\rm min}$ (middle panels), and $\gcutminratio$ (bottom panels). The left, middle, and right columns refer to the selected values of $q$, equal to 0, 1, and -1.5, respectively. The dashed horizontal lines mark the    0.05 $p$-value.}
\label{fig:fig_full_mc_vs_data_robopol}
\end{figure}

\subsection{RoboPol limiting envelope}
To test whether the samples generated in our MC are statistically compatible with the RoboPol observed trend, we use a two-dimensional two-sample Kolmogorov–Smirnov (KS) test, where the two independent variables are $\nu_p^S$ and $\Pi_{\rm opt}$ (evaluated at $5\times 10^{14}$ Hz). The KS test is performed using the public code \textsc{ndtest} \footnote{Written by Zhaozhou Li, \url{https://github.com/syrte/ndtest}}, and crosschecked against the public code \textsc{2DKS}  \footnote{Written by Gabriel Taillon, \url{https://github.com/Gabinou/2DKS}, \citep{Peacock1983,Fasano1987,Press2007}}.
In Figure \ref{fig:fig_full_mc_vs_data_robopol} we show the trends of the KS test $p$-value as a function of $\Delta_p$ (top panels), $r_{\rm min}$ (middle panels) and $\gcutminratio$ (bottom panels).

\begin{figure}
   \centering
   \includegraphics[width=\columnwidth]{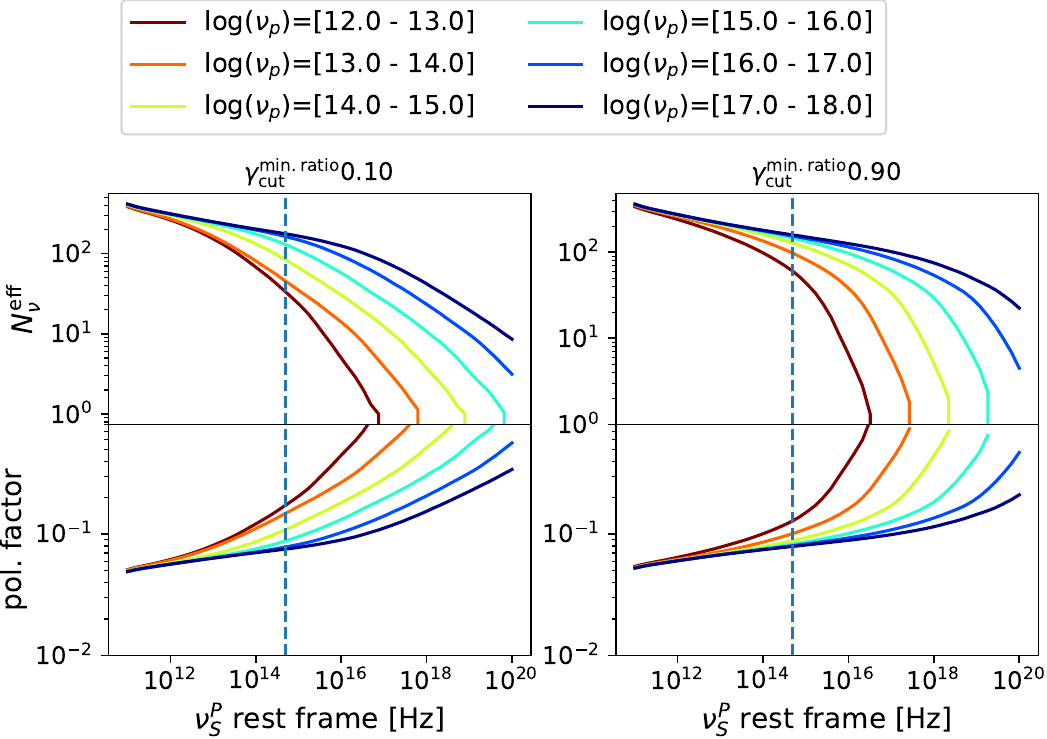}\\
\caption{Trial-averaged trends for $N_\nu^{\rm eff}$ and  for the polarization factor, $\kappa_\nu^{\rm eff}$, for the  PL scenario with $f_{p_c}=\mathcal{U}[1.5,2.5]$, $q=-1.0$, $r_{\rm min}=0.1$, and  $\fgammacut$=\emph{log-uniform}. The different values of $\gcutminratio$  are reported in the panel title, and the dashed vertical line marks the reference value for the optical frequency.}
\label{fig:robopol_gmin_ratio_trends_1}
\end{figure}
\begin{figure}
   \centering
   \includegraphics[width=.9\columnwidth]{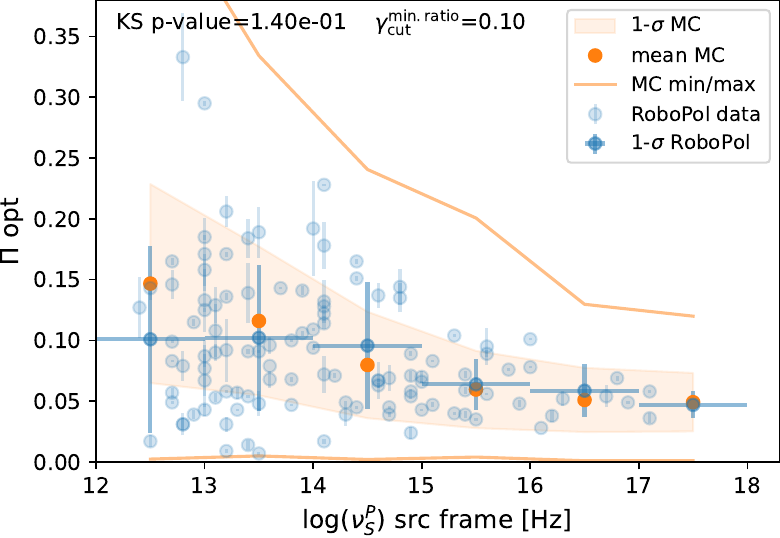}\\
   \vspace{.2cm}
   \includegraphics[width=0.9\columnwidth]{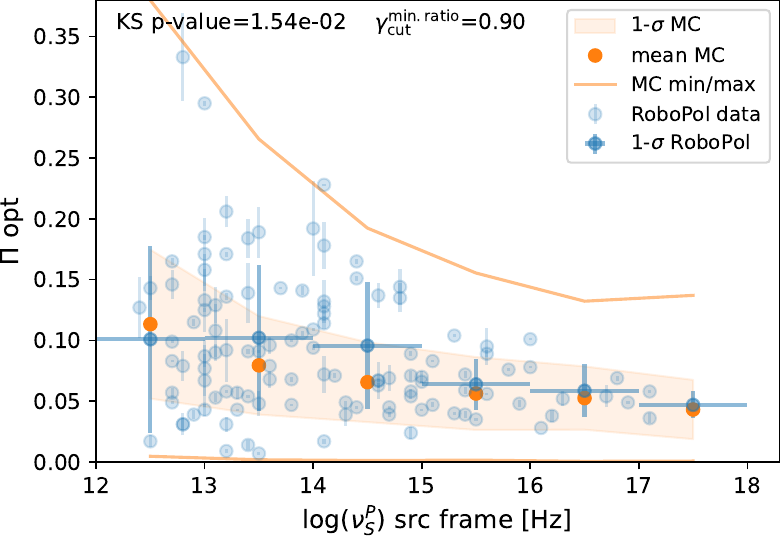}\\
  
   \caption{MC $\Pi_{\rm opt}$-versus-$\nu_p^S$  trend, compared to the RoboPol observed one, for the same MC run configuration as reported in Figure \ref{fig:robopol_gmin_ratio_trends_1}.
   The panels from top to bottom show the trend for increasing values of $\gcutminratio$, from $\gcutminratio=0.01$  (top panel) to $\gcutminratio=0.9$ (bottom panels). 
   The $\Pi_{\rm opt}$ trend flattens for larger values of $\gcutminratio$, with the KS $p$-value decreasing from $\approx 0.14$, for the case of $\gcutminratio=0.1$, to a $p$-value  $\approx 0.015$, for the case of $\gcutminratio=0.9$}
\label{fig:robopol_gmin_ratio_trends_2}
\end{figure}

\begin{figure*}
   \centering
   \includegraphics[width=0.9\textwidth]{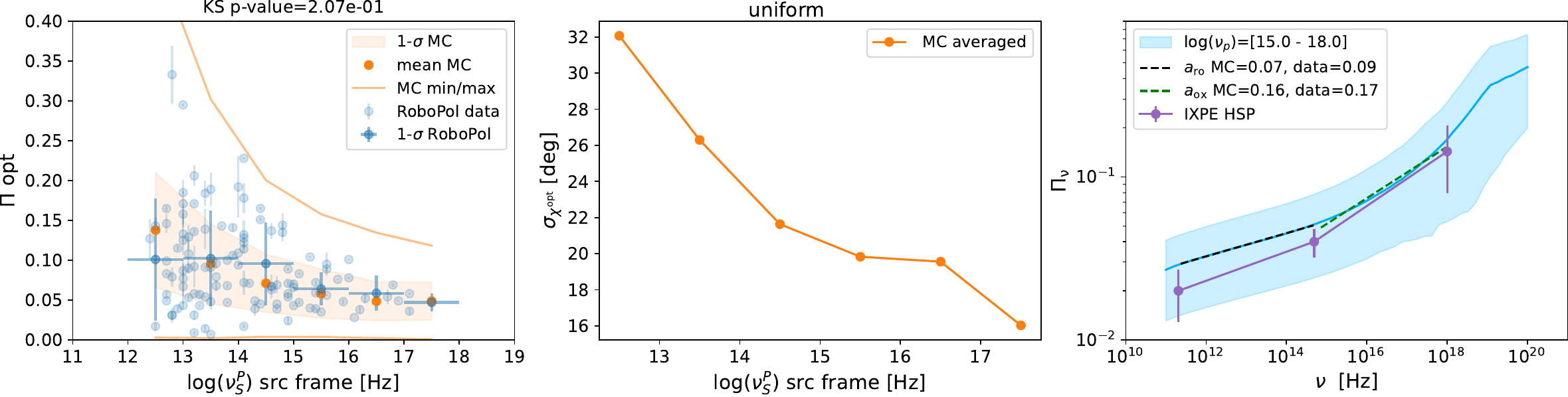}\\
   \includegraphics[width=0.9\textwidth]{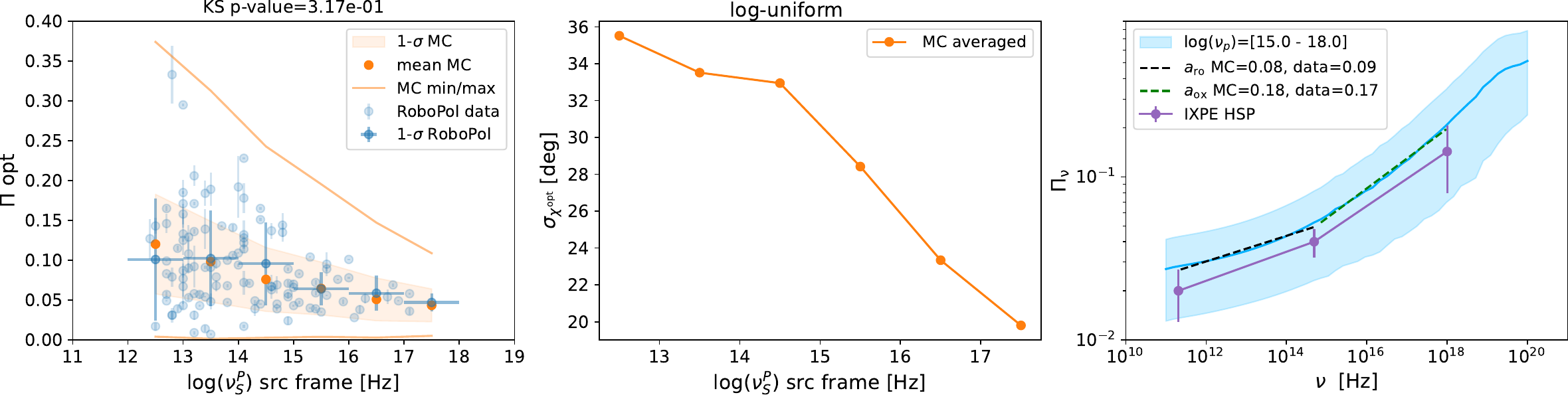}\\
   \includegraphics[width=0.9\textwidth]{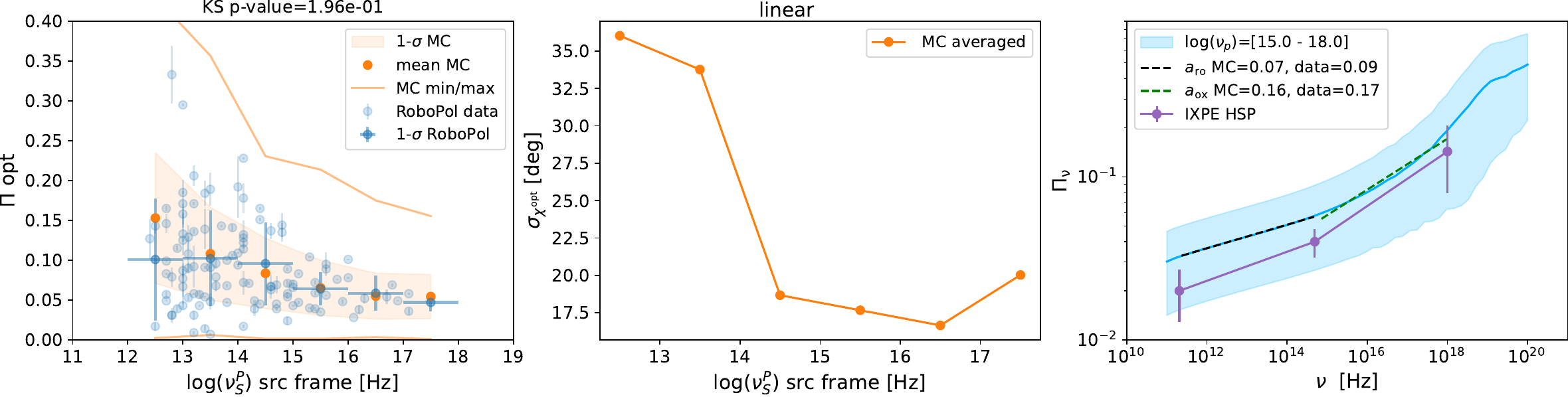}\\
   \caption{Best runs for $\fgammacut$= \emph{uniform} (top panels), $\fgammacut$= \emph{log-uniform} (middle panels), and $\fgammacut$= \emph{linear} (bottom panels). 
The left panels show the $\Pi_{\rm opt}$-versus-$\nu_p^S$ limiting envelopes for both the MC simulations and the observed RoboPol data, the center panels show the trends of the MC-averaged dispersion of the MC optical EVPA angle vs $\nu_p^S$, and in the right panels, the IXPE HPS observed trends compared to the MC results extracted from the same runs used for the RoboPol panels.}
   \label{fig:pol_full_mcpol_full_mc_vs_data_delta_chi_opt}
\end{figure*}

We notice that consistently with the HSPs IXPE trends, the RoboPol  $\Pi_{\rm opt}$-versus-$\nu_p^S$ limiting envelope rules out the $\fgammacut$=\emph{fixed} PDF, and disfavor values of $\gcutminratio\gtrapprox 0.1$, for all the $\fgammacut$ PDFs, and values of $r_{\rm min}\gtrapprox 0.1$ for the \emph{linear} $\fgammacut$. 
Anyhow, the $\Pi_{\rm opt}$-versus-$\nu_p^S$ limiting envelope does not constrain the dispersion on $\Delta_p$. The reason for this behavior stems from the fact that the observed RoboPol $\Pi_{\rm opt}$-versus-$\nu_p^S$  envelope depends mostly on the `distance' between the optical frequency and $\nu_p^S$, and in particular on the modulation of $N_\nu^{\rm eff}$, that is, the polarization factor $\kappa_\nu^{\rm eff}$  below $\nu_p^S$. 
This effect is shown in Figures \ref{fig:robopol_gmin_ratio_trends_1} and \ref{fig:robopol_gmin_ratio_trends_2}. In Figure \ref{fig:robopol_gmin_ratio_trends_1} we plot the trial-averaged trends for $N_\nu^{\rm eff}$ and for the polarization factor, $\kappa_\nu^{\rm eff}$ for the PL scenario with $f_{p_c}=\mathcal{U}[1.5,2.5]$, $q=-1.0$, $r_{\rm min}=0.1$, and  $\fgammacut$=\emph{log-uniform}. 
The different values of $\gcutminratio$ are reported in the panel title, and the dashed vertical line marks the reference value for the optical frequency. 
We notice how increasing  $\gcutminratio$ from 0.1 to 0.9, the optical excursion across the $\nu_p$ range, of both  $N_\nu^{\rm eff}$ and $\kappa_\nu^{\rm eff}$, decreases. 
This leads to a different trend in the $\Pi_{\rm opt}$-versus-$\nu_p^S$ plane, as shown in Figure  \ref{fig:robopol_gmin_ratio_trends_2}, where we notice the increase of $\gcutminratio$ from 0.1 to 0.9  leads to a flattening of the limiting envelope, with the KS $p$-value decreasing from $\approx 0.14$, for the case of $\gcutminratio=0.1$, to a $p$-value $\approx 0.015$, for the case of $\gcutminratio=0.9$.
Finally, in Figure \ref{fig:pol_full_mcpol_full_mc_vs_data_delta_chi_opt} we plot the best runs for $\fgammacut$=\emph{uniform} (top panels), $\fgammacut$=\emph{log-uniform} (middle panels), and $\fgammacut$=\emph{linear} (bottom panels). The left panels show the   $\Pi_{\rm opt}$-versus-$\nu_p^S$ limiting envelopes for both the MC simulations and the observed RoboPol data, the center panels show the dispersion trends of the MC optical EVPA angle, $\sigma_{\chi}^{\rm opt}$ vs $\nu_p^S$, and the right panels show the IXPE HSP observed $\Pinuave$ trends compared to the MC results, extracted from the same runs used for the RoboPol panels. 
The decreasing trend of the optical EVPA angle dispersion has the same root as the $\Pi_{\rm opt}$-versus-$\nu_p^S$. i.e. the modulation of  $N_\nu^{\rm eff}$ below $\nu_p^S$. For lower values of $\nu_p^S$, the optical $N_\nu^{\rm eff}$ decreases; hence, the dispersion on the optical EVPA angle increases. 

\subsection{Phenomenology constraint on the MC parameter space}
\begin{figure}
   
   \centering   
   \includegraphics[width=0.35\textwidth]{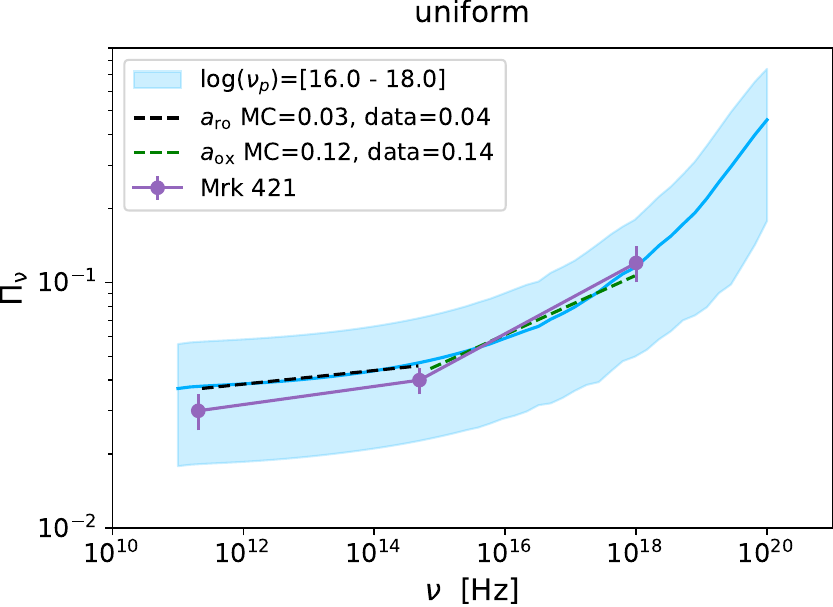}\\    
   \includegraphics[width=0.35\textwidth]{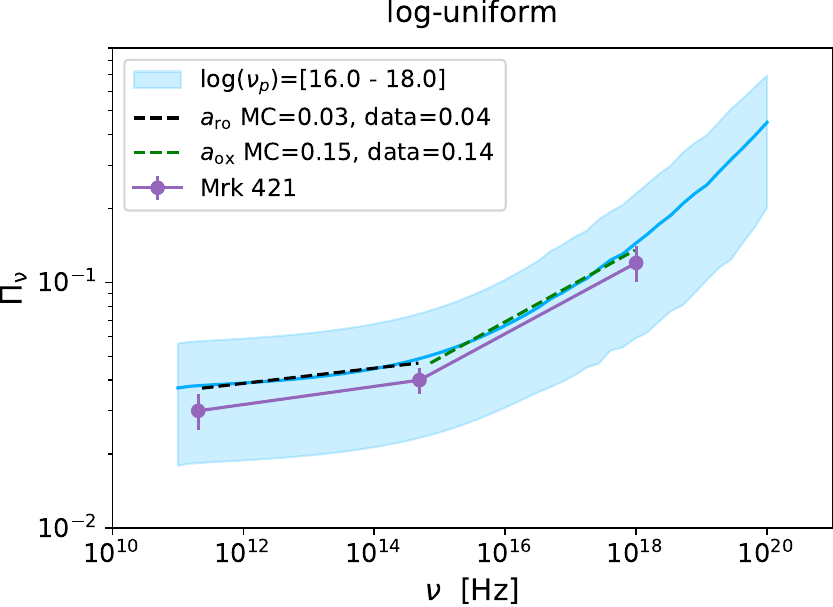}\\   
   \includegraphics[width=0.35\textwidth]{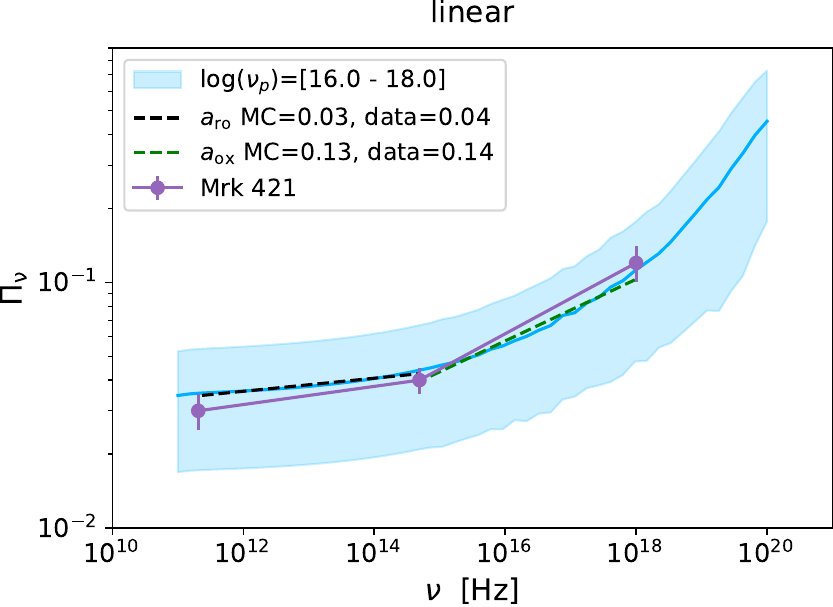}   
   \caption{$\Pi_{\nu}$ trends for the top-ranked MC runs tested against the observed data of Mrk 421 for the case of  $\fgammacut$= \emph{uniform} (top panel), $\fgammacut$= \emph{log-uniform} (middle panel), and $\fgammacut$= \emph{linear}  (bottom panel).}
   \label{fig:test_case_chromaticity}
\end{figure}

\begin{figure*}
   \centering
   \includegraphics[width=0.75\textwidth]{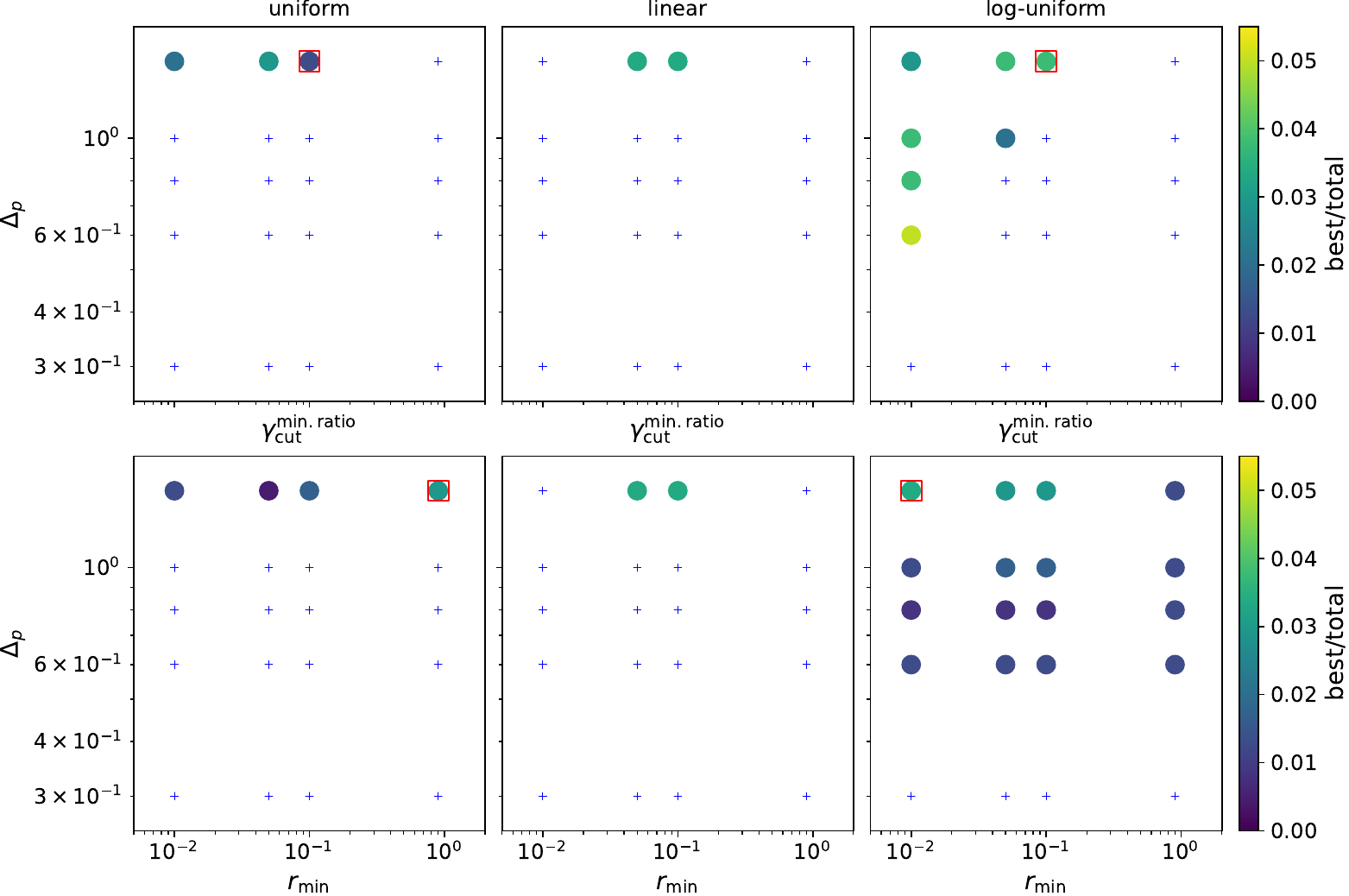}
   \caption{Statistical outcome of the PL-distributed MC parameter space selection. 
The  $\Delta_p$-versus-$\gcutminratio$ parameter space is shown in the top panels, whilst the bottom panels show the  $\Delta_p$-versus-$r_{\rm min}$ parameter space. The three columns refer to the selected $\fgammacut$ PDF. The blue crosses mark the total  MC parameter space, and the filled circles the \emph{best} sample parameter space. 
The color scale marks the ratio of the volume of \emph{best} parameter space, at the coordinate reported on the x and y axes, to the total MC parameter space; both volume sizes refer to the given $\fgammacut$ PDF. The red boxes mark the parameter space where the MC-averaged $1$-$\sigma$ dispersion of the X-ray EVPA is  $<$25\textdegree.}
   \label{fig:full_mc_best_par_space}
\end{figure*}

The two tests against IXPE and RoboPol data have provided an interesting constraint on the MC parameter space.
We define our \emph{best} sample by selecting runs resulting in a  RoboPol KS $p$-value $>0.1$,  and a $20\%$ maximal relative difference between the IXPE and MC for both the  $a_{\rm mm-o}$ and $ a_{\rm o-X}$.
The selection reduces the volume of the parameter space, from 780 (for the total MC volume) to 79, for the best selection volume.
In Figure \ref{fig:full_mc_best_par_space} we show the statistical outcome of the selection. 
In the top panels, we report the $\Delta_p$-versus-$\gcutminratio$, and in the bottom panels the  $\Delta_p$-versus-$r_{\rm min}$ parameter space. 
The three columns refer to the selected $\fgammacut$ PDF. 
The blue crosses mark the total  MC parameter space, and the filled circles the \emph{best} sample parameter space. 
The color scale marks the ratio of the volume of \emph{best} parameter space, at the coordinate reported on the x and y axes, to the total MC parameter space; both the volume sizes refer to the given $\fgammacut$ PDF.
The plots confirm the results discussed in the previous section, i.e., the observed and combined RoboPol/IXPE phenomenology constrain $\Delta_p\gtrapprox 0.6$, and $\gcutminratio\lessapprox 0.1$.
The \emph{log-uniform} $\fgammacut$ PDF covers a larger volume of the parameter space, and reaches the peak value of the fractional acceptance, at almost 0.05. 
We find it quite promising that by only imposing the MC calibration condition of $\Pi_{\rm opt}\approx 0.045$, for HSP sources, our model was able to reproduce simultaneously both the IXPE and RoboPol observed phenomenology.
Finally, we test our MC-predicted $\Pi_\nu$ trends against the data of Mrk~421, showing pronounced chromaticity in $\Pi_\nu$. To this end, we restrict the $\nu_p^S$ bin to $[10^{16},10^{18}]~\rm Hz$ (consistent with the source SED) and select the runs that best match the observed values of $a_{\rm mm-o}$ and $a_{\rm o-X}$, ranking them by the $\chi^2$ computed from each run’s median (0.5-quantile) predictions.
In Figure~\ref{fig:test_case_chromaticity} we show the top-ranked runs for $\fgammacut$=\emph{uniform} (top panel), \emph{log-uniform} (middle panel), and \emph{linear} (bottom panel). The MC $\Pi_\nu$ trends reproduce the observed chromaticity of Mrk~421 with good accuracy. This contrasts with \cite{Liodakis2022}, who argued that multi-zone models can reproduce only a mild level of chromaticity in $\Pi_\nu$.

\section{Discussion and conclusions}
\label{sec:disc_concl}

We have presented an analysis of the multiwavelength pattern in the $\Pi_\nu$ of blazars, based on a comprehensive Monte Carlo framework modeling multi-zone synchrotron emission, able to reproduce the energy-dependent polarization characteristics observed from millimeter to X-ray wavelengths. 
Our approach implements spatially resolved emitting regions, with distinct physical properties, in a purely turbulent framework, that is, we do not assume any correlation between magnetic field and cell size and/or position; we only assume a power-law distribution for the cell size, and we test different levels of correlation between the size of the cells, and cutoff in the electron distribution. 
This minimal approach aims to provide the most likely statistical characterization of the cell distributions, in terms of size and energy of the electrons, capable to reproduce the energy-dependent polarization patterns observed in the millimiter-to-X-ray data for IXPE HPS, and in the optical for the RoboPol dataset. Our main findings are summarized below:

\begin{itemize}

   \item The frequency-dependent polarization fraction, $\Pi_\nu$ does not actually depend on the number of emitting cells contributing to a given frequency, $N_\nu$,  but on the flux-weighted average effective number $N_{\nu}^{\rm eff}$, which gives the actual contribution from each cell. This effect plays a relevant role when the size of the cells and the EED have significant dispersions. 
   According to the distributions of the cell size, and/or of the EED parameters, the difference between  $N_\nu$ and $N_{\nu}^{\rm eff}$ can be of a few orders of magnitude, with $N_{\nu}^{\rm eff}<N_\nu$.
   Hence, using the estimate of $N_\nu$, for example, from X-ray measurements of $\Pi_\nu$, leads to an underestimation of the actual number, $N_c$, of cells populating the region, and in the number of cells radiating at lower frequencies.
   This effect is significant both for the total and the \emph{best} MC parameter space (see Figure \ref{fig:N_c_to_N_eff}).
  
   \item In general, $\Pi_\nu$ shows a broken power-law shape, with the turnover frequency around $\nu_p^S$. 
   For the case of no dispersion on the low-energy index of the EED,  $\Delta_p=0$, the low energy spectral index $a_l\approx 0$, whilst it increases as $\Delta_p$ increases, up to a value of $a_l\approx 0.1$ for  $\Delta_p=1.5$. $a_l$ results to be in general independent on $q$, $r_{\rm min}$, and $\gcutminratio$. 
   The only $\fgammacut$ distribution showing a dependence on $\gcutminratio$ is the  \emph{log-uniform}, with $a_l$ decreasing as $\gcutminratio$ increases. 
   This is expected since $\fgammacut\propto \gamma_{\rm cut}^{-1}$ and the cell size is uncorrelated with $\gamma_{\rm cut}$; thus, lower values of $\gcutminratio$ lead to a larger dispersion in the cell flux weights at low frequencies, increasing $N_{\nu}^{\rm ef}$, with a consequent hardening in the slope of the polarization factor below $\nu_p^S$. 
   The high-energy slope, $a_h$, of the polarization fraction, unlike $a_l$, is independent of $\Delta_p$,  and generally also of $r_{\rm min}$ and $\gcutminratio$, since the modulation of $N_{\nu}^{\rm eff}$ is mainly dictated by the cell flux contributions above $\nu_p$, which depends on the dispersion above $\gamma_{\rm cut}$.

   \item We have tested our MC  $\Pi_\nu$ trends against the IXPE HPS multiwavelength datasets. We find that the observed $a_{\rm mm-o}$ index significantly constrains the dispersion on $p$, favoring a dispersion $\Delta_{p}\gtrapprox 1$, and values of $\gcutminratio\lessapprox 0.1$ for the \emph{log-uniform} distribution. 
   We also noticed that \emph{log-uniform} distribution provides the best match with the data, and that the scenario with a fixed cutoff in the EED is ruled out. The high-energy index, $a_{\rm o-X}$, provides a lower constraining power, anyhow, a significant tension with the data is observed for $r_{\rm min}\approx 1$, for the case of \emph{linear} $\fgammacut$, and for values of $\gcutminratio\approx 1$, for all the   $\fgammacut$ distributions.

   \item The result from the test of the  MC  $\Pi_{\rm opt}$-versus-$\nu_p^S$ limiting envelope against the observed RoboPol one is consistent with the outcome of the test against the IXPE HPS multiwavelength datasets. RoboPol data rules of the scenario with a fixed cutoff in the EED. Values of $\gcutminratio \gtrsim 0.1$ are disfavored for all tested $\fgammacut$ PDFs.
   The trend is primarily sensitive to the modulation of the effective number of emitting cells at optical frequencies, which depends on the distance between the optical band and the synchrotron peak. 
   The observed limiting envelope between optical polarization and $\nu_p^S$ is well reproduced for models with a broad distribution of cutoff energies and a sufficiently large dispersion in the EED index $p$. Interestingly, we find that the same driver of the  $\Pi_{\rm opt}$-versus-$\nu_p^S$ leads to a similar trend for the dispersion of the MC optical EVPA angle, $\sigma_{\chi}^{\rm opt}$ vs $\nu_p^S$, that is, an anticorrelation with $\nu_p^S$, in agreement with the observed RoboPol one.

   \item The test of our simulations against combined IXPE and RoboPol observational data has provided a further constraint on model parameter space. By selecting MC runs that simultaneously match the RoboPol limiting envelope (KS $p$-value $>0.1$) and reproduce the IXPE millimiter-to-optical and optical-to-X-ray polarization slopes within 20\% of the observed values, the allowed parameter space (reported in Figure \ref{fig:full_mc_best_par_space}) is reduced by a factor of ten, compared to the full MC parameter space. 
   The analysis shows that the best agreement is obtained for models with a broad dispersion in the EED index ($\Delta_p \gtrsim 0.6$) and values of $\gcutminratio \lesssim 0.1$ (that is, a large dispersion on $\gamma_{\rm cut}$), with the  \emph{log-uniform} $\fgammacut$ distribution providing the best agreement. 
\end{itemize}

The constraints discussed above hint to both a wide range of single-cell electron spectral indices and a broad distribution of single-cell cutoff energies to reproduce the observed S multiwavelength polarization properties of blazars.
The results demonstrate that a minimal, turbulence-driven multi-zone scenario can account for the key features seen in both IXPE and RoboPol datasets.
We also notice that the \emph{linear} model does not provide statistical improvement compared to the \emph{log-uniform}, hence, the current dataset and analysis are not able to find significant evidence for this model, which mimics statistically the magnetic reconnection scenario. Nevertheless, this scenario is not ruled out. 

Our results regarding the IXPE $\Pi_\nu$ trends are consistent with those presented in \cite{Peirson2019}; however, we have investigated a broader parameter space and presented a quantitative prediction of the related phenomenology, validated by a test against recent observational data from millimiter to X-ray frequencies. Moreover, we notice that contrary to what is reported in \cite{Liodakis2022}, the multi-zone scenario is capable of reproducing the observed chromaticity of the $\Pi_\nu$ trends.
We stress that our model does not test the presence or not of a shock, and that the investigation on the dispersion of the optical EVPA is mostly limited to the multi-zone scenario, with any implication on observed systematic rotations at different wavelengths. 
Hence, our analysis can fit a purely stochastic scenario or a scenario where the turbulent medium develops within the shock. In this regard, we notice how HSP objects, such as Mrk 421, have shown in their spectral shapes evidence for the coexistence of both first-order shock acceleration and stochastic acceleration \citep{Tramacere2009, Tramacere2011}. 
The presence of a shock is supported by recent population studies of IXPE results \citep{Chen2024,Capecchiacci2025} providing an important observational constraint: they report EVPA–jet alignment in X-rays with a dispersion on the order of 20\textdegree (except for the larger rotations seen in Mrk~421  \citep{DiGesu2023}). If the EVPA–jet stable alignment reflects shock acceleration with in-shock turbulence, the MC EVPA should exhibit low dispersion, particularly in X-rays, in agreement with the IXPE results.
For this purpose, in Figure \ref{fig:full_mc_best_par_space} we marked with red boxes the points in our MC \emph{best} parameter space that yield, on average, a $1$-$\sigma$ dispersion of the X-ray EVPA below 25\textdegree. This selection restricts the best parameter space to $\fgammacut$ distributed \emph{uniformly} or \emph{log-uniformly}, with $\Delta_p=1.5$. In particular, for the $\fgammacut$= \emph{uniform} case we obtain constraints $r_{\rm min}=0.9$ and $\gcutminratio=0.1$, while for the $\fgammacut$= emph{log-uniform} case the constraints are $r_{\rm min}=0.01$ and $\gcutminratio=0.1$. Across the MC parameter space, the  MC-averaged $1$-$\sigma$ dispersion of the X-ray EVPA is $<$40\textdegree, and roughly $75\%$ of the \emph{best} region is compatible with a dispersion $<$35\textdegree. These constraints can help to clarify the turbulence structure developed in the jet, during states when jet-axis/EVPA alignment is observed, but stronger conclusions will require more comprehensive shock-plus-turbulence modeling and longer, more densely sampled X-ray polarimetric observations of the sources.

\begin{acknowledgements}
A.T. would like to thank the anonymous referee for carefully reading the manuscript and for giving constructive comments, which helped to improve the quality of the paper,
and Enrico Massaro for fruitful discussions.
\end{acknowledgements}

\bibliographystyle{aa}
\bibliography{stoch_pol} 

\begin{appendix}
\section{MC calibration stage}
\label{sec:calibration_stage}
The  calibration stage, performed for each run,  calibrates $N_c$ and 
$\gamma_{\rm cut}^{\rm ref}$, to obtain a trial-averaged optical polarization fraction, $\Pinuaveopt$, within $[0.4-0.5]\%$, and a trial-averaged value of $\nu_p^S$ within=$[0.8-1.2] \times 10^{17}$ Hz. As a reference value for the optical frequency, we use the value of $5\times 10^{14}$ Hz. The calibration stage proceeds as follows:
\begin{enumerate}

   \item Initially, we set $N_c=1000$, and we set $\gamma_{\rm cut}^{\rm calib}$ by a numerical sampling.
   In detail, we define a range of 100 $\gamma_{\rm cut}^{\rm calib}$ values, evenly spaced in the logarithm, in the range $[10,10^8]$, and for each value we estimate the value of $\nu_p^S$, according to the third of Eq.\ref{eq:delta_approx_n_gamma}, plugging the expectation values of $\overline{f_B}$, $\overline{f_\delta}$, and  $\overline{f_p}$. The value of $\gamma_{\rm cut}^{\rm calib}$ resulting in the best match to $\nu_p^S=1 \times 10^{17}$ Hz is chosen.
   
   \item A first 10-trial run is performed with the value of  $\gamma_{\rm cut}^{\rm calib}$ obtained at the previous step. 
   The resulting trials-averaged values of $\nupSave$ and the $\Pinuaveopt$ are tested against the reference values,  within a relative tolerance of 0.02. If the difference exceeds the tolerance, the previous step is repeated updating the value of $N_c$ according to equation \ref{eq:frac_pol_same_cells}, and the value of  $\gamma_{\rm cut}^{\rm calib}$ is updated compensating the relative offset on $\nupSave$ according to the proportionality established by the third of, Eq.\ref{eq:delta_approx_n_gamma}.

\end{enumerate}

\section{N cells versus N eff}
\begin{figure*}
\centering
\includegraphics[width=.8\columnwidth]{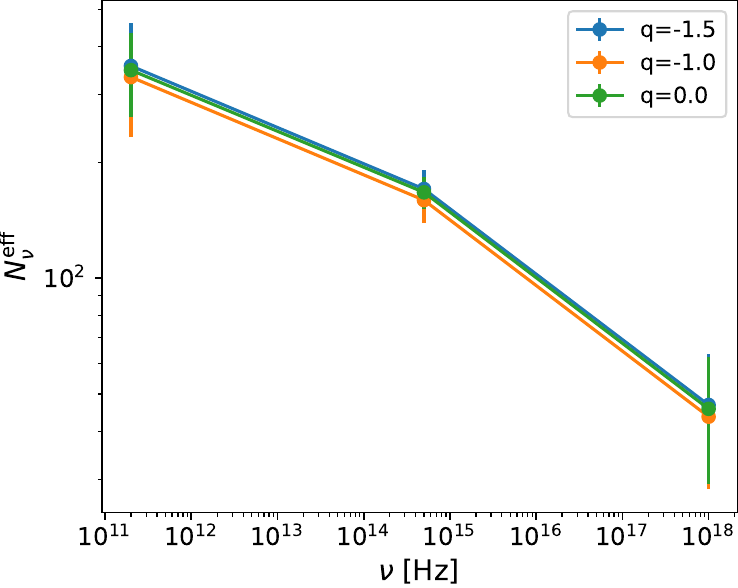}
\includegraphics[width=.8\columnwidth]{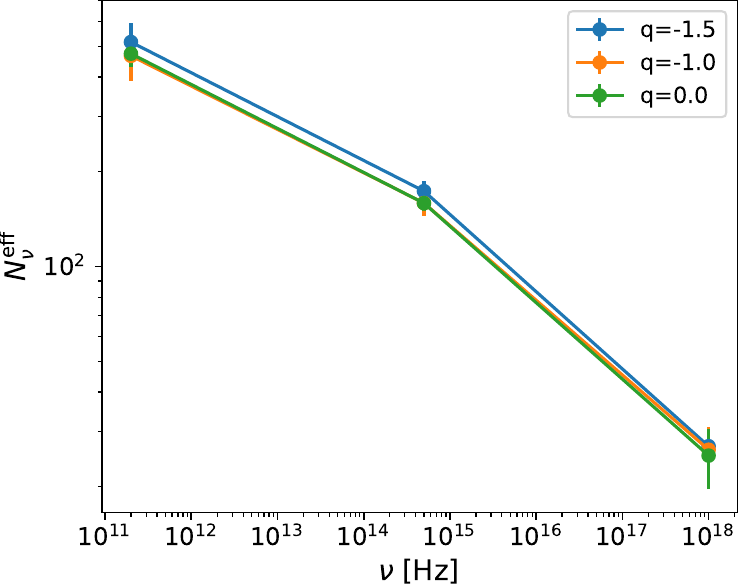}
\caption{The trend of the effective flux-averaged, $N_{\rm \nu}^{\rm eff}$, at the reference frequencies used for the  
millimiter ($2\times 10^{11}$  Hz), optical ($5\times 10^{14}$ Hz),  and X-ray ($1\times 10^{18}$ Hz) frequencies. 
Different color lines mark different values of $q$ as detailed in the legend.
The left panel refers to the full PL MC parameter space, the right panel to the \emph{best} sample parameter space.}
\label{fig:N_eff_vs_nu}
\end{figure*}

\begin{figure*}
\centering
\includegraphics[width=.8\textwidth]{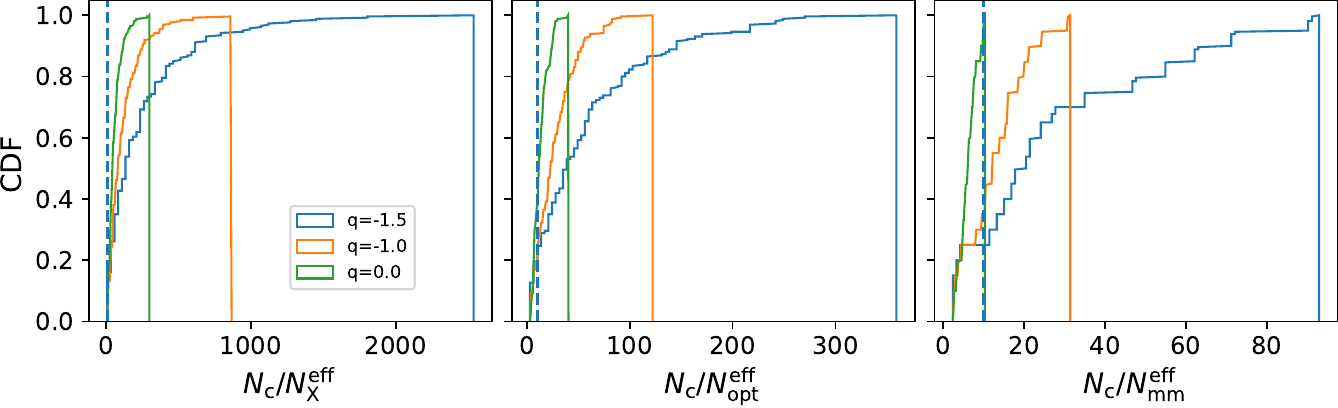}\\
\includegraphics[width=.8\textwidth]{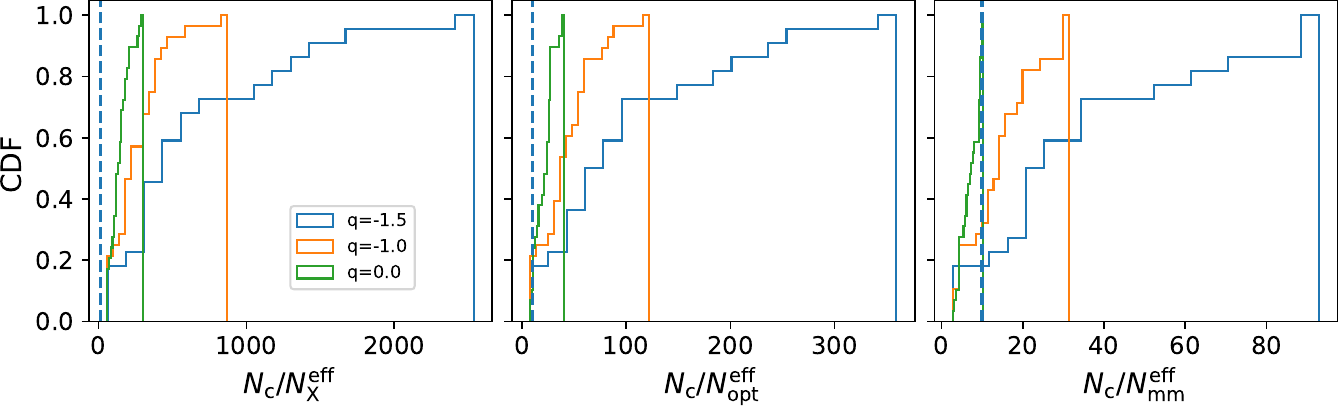}
\caption{The cumulative distribution function (CDF) for the ratio of the total number of cells, 
$N_c$, to the value of the effective flux-averaged, $N_{\rm \nu}^{\rm eff}$, at the reference frequencies used for the  
millimiter ($2\times 10^{11}$  Hz), optical ($5\times 10^{14}$ Hz),  and X-ray ($1\times 10^{18}$ Hz) frequencies, 
shown respectively in the left, center, and right panels. 
Different color lines mark different values of $q$ as detailed in the legend.
The top panels refer to the full PL MC parameter space, the bottom panels to the \emph{best} sample 
parameter space. The  vertical dashed line marks $N_c/N_{\rm \nu}^{\rm eff}=10$.}
\label{fig:N_c_to_N_eff}
\end{figure*}

We investigate the dependence of $N_c/N_{\nu}^{\rm eff}$ for different values of $q$.
In Figure \ref{fig:N_eff_vs_nu} we show the MC trends for the effective flux-averaged, $N_{\rm \nu}^{\rm eff}$, at the reference frequencies used for the  millimiter ($2\times 10^{11}$  Hz), optical ($5\times 10^{14}$ Hz),  and X-ray ($1\times 10^{18}$ Hz) frequencies. 
The dependence of $N_c/N_{\nu}^{\rm eff}$ for difference values of $q$ is reported in \ref{fig:N_c_to_N_eff},  where the top panels refer to the full PL MC parameter space, and the bottom panels to the \emph{best} sample parameter space. We notice how the number of $N_c$ exceeds by at least a factor 
of 10  the value of $N_{\nu}^{\rm eff}$. The upper value of $N_c/N_{\nu}^{\rm eff}$ reaches up to a few thousands, 
for the X-ray ($\nu=10^{18}$ Hz) frequency, and decrease to a few hundreds for the optical frequency ($\nu=5\times 10^{14}$ Hz),
and up to a few tens for millimiter frequency ($\nu=2\times 10^{1}$ Hz). The effect is larger for steeper values of $q$, 
meaning that a larger dispersion on the cell size amplifies the difference between $N_c$ and $N_{\nu}^{\rm eff}$, 
due to the higher impact on the flux-weighted average effective number $N_{\nu}^{\rm eff}$.

\end{appendix}
\end{document}